# Phosphorene: Synthesis, Scale-up, and Quantitative Optical Spectroscopy

*Adam H. Woomer[§#], Tyler W. Farnsworth[§#], Jun Hu[§], Rebekah A. Wells[§], Carrie L. Donley[‡], and Scott C. Warren[§†]\**

[§]Department of Chemistry, [‡]Chapel Hill Analytical and Nanofabrication Laboratory, and [†]Department of Applied Physical Sciences, University of North Carolina at Chapel Hill, Chapel Hill, NC 27599, USA.

[#]These authors contributed equally.

KEYWORDS 2D materials, black phosphorus, liquid exfoliation, quantum confinement, band gap

ABSTRACT Phosphorene, a two-dimensional (2D) monolayer of black phosphorus, has attracted considerable theoretical interest, although the experimental realization of monolayer, bilayer, and few-layer flakes has been a significant challenge. Here we systematically survey conditions for liquid exfoliation to achieve the first large-scale production of monolayer, bilayer, and few-layer phosphorus, with exfoliation demonstrated at the 10-gram scale. We describe a rapid approach for quantifying the thickness of 2D phosphorus and show that monolayer and few-layer flakes produced by our approach are crystalline and unoxidized, while air exposure leads to rapid oxidation and the production of acid. With large quantities of 2D phosphorus now available, we perform the first quantitative measurements of the material's absorption edge—which is nearly identical to the material's band gap under our experimental conditions—as a function of flake thickness. Our interpretation of the absorbance spectrum relies on an analytical method introduced in this work, allowing the accurate determination of the absorption edge in polydisperse samples of quantum-confined semiconductors. Using this method, we found that the band gap of black phosphorus increased from $0.33 \pm 0.02$ eV in bulk to $1.88 \pm 0.24$ eV in bilayers, a range that is larger than any other 2D material. In addition, we quantified a higher-energy optical transition (VB-1 to CB), which changes from 2.0 eV in bulk to 3.23 eV in bilayers. This work describes several methods for producing and analyzing 2D phosphorus while also yielding a class of 2D materials with unprecedented optoelectronic properties.

KEYWORDS Phosphorene, black phosphorus, liquid exfoliation, band gap, quantum-confined, 2D materials, optical spectroscopy

Solution-processable nanomaterials with tunable optoelectronic properties are being considered as potential building blocks for numerous technologies, such as photovoltaics,[1] transistors,[2] and light-emitting diodes.[3] Among these nanomaterials, quantum dots have attracted broad interest because of their size-dependent electronic structure and controllable physical properties; for example, band gaps can be increased by as much as 2 eV as particle size decreases.[4–7] With the advent of two-dimensional (2D) semiconductors,[8] new opportunities have emerged for designing materials and devices, although the size-dependent variation of electronic properties like band gaps are, in general, smaller: transition metal dichalcogenides have band gaps that can only be tuned by 0.7 eV[9,10] while, for example, PbSe quantum dots can be tuned from 0.27 to 1.5 eV.[11–13] Toward increasing the library of solution-processable materials, here we show that black phosphorus can be liquid exfoliated to yield a family of 2D flakes with tunable optical properties that rival those of quantum dots.

Black phosphorus,[14] a layered 3D crystal of elemental phosphorus (Figure 1a), and its 2D derivative, termed phosphorene[15,16] (Figure 1b), have recently attracted renewed[17] attention. In the last few months, there have been exciting demonstrations of the material's application to transistors,[16,18,19] photovoltaics,[20,21] photodetectors,[22,23] and batteries.[24,25] As a 2D material with an intriguing corrugated or accordion-like structure, phosphorene has captured significant theoretical interest with numerous predictions of the material's anisotropic[16,26] and thickness-dependent optoelectronic properties,[27,28] mechanical properties,[29] and chemical reactivity.[30–32] Most predictions have gone untested, however, because there is still no reliable method to make or purify monolayer or few-layer phosphorus. When monolayers have been observed, they are typically situated at the edges of thicker sheets and are typically too small to characterize. Underlying these practical challenges are the inherent problems associated with phosphorus: the phosphorus-phosphorus bonds are significantly weaker than carbon-carbon bonds and several studies have noted the material's tendency to oxidize[14,33] or form other allotropes.[34,35] In addition, interlayer interactions may be stronger in black phosphorus than in other 2D materials.[36,37] These strong interlayer interactions would inhibit exfoliation and, consequently, black phosphorus may be harder to exfoliate and more likely to fragment than other 2D materials. In fact, this is consistent with reports of mechanical exfoliation in which sheets of fewer than six layers have seldom been observed.[16,18,19,38,39]

Our own attempts to mechanically exfoliate black phosphorus confirmed the results of other groups. We prepared and analyzed samples under an inert atmosphere, using scotch tape for exfoliation and a Bruker Dimension FastScan atomic force microscope (AFM) to rapidly analyze sheet thickness over macroscopic areas (see Supporting Information for additional details). We randomly surveyed large areas and assessed the structure of over 3,000 flakes. Our survey revealed that the yield of sheets thinner than 10 layers is less than 0.06%; in addition, no sheets thinner than 6 layers were found. Given the low odds for identifying and characterizing 2D materials prepared in this way, we began exploring liquid exfoliation[40,41] as an alternative route for material preparation. Here we provide evidence that liquid exfoliation, when carefully executed under an inert atmosphere, produces macroscopic (milligram-to-gram scale) quantities of monolayer and few-layer phosphorene.[42] We note that this is a considerable improvement over state-of-the-art methods of liquid exfoliation,[43–45] which have so-far produced flakes with thicknesses that are 10 to 20 times thicker than those described here. We characterize the material's structure, stability, and thickness-dependent optical properties and compare these properties to theoretical predictions. In addition, we perform the first quantitative optical absorption measurements on 2D phosphorus, allowing us to determine the thickness-dependent optical transitions and band gaps.

## RESULTS AND DISCUSSION

### *Liquid exfoliation of black phosphorus*

Black phosphorus crystals (Figure 1a) were acquired from Smart Elements between December 2012 and March 2014 or grown in our laboratory by $SnI_2$ vapor transport.[46] (Smart Elements modified its method of manufacture in the summer of 2014 and the microstructures of materials acquired after this date may differ.) Black phosphorus was ground in a mortar and pestle and sonicated in anhydrous, deoxygenated organic liquids using low-power bath sonication under an inert atmosphere. In our initial experiments, black phosphorus was sonicated in electronic grade isopropanol for sixteen hours. During sonication, the phosphorus was suspended in solution and its color changed from black to reddish-brown to yellow (Figure 1c), indicating a profound change in the electronic structure of the material. We quantified this change in appearance by ultraviolet-visible-near IR (UV-vis-near IR) absorption spectroscopy (see discussion below for further details). Over several weeks, there was limited reaggregation and no further change in color, suggesting that these suspensions were comprised of small phosphorus particulates. To examine the morphology of the particulates, suspensions were drop-cast onto a silicon wafer for analysis by scanning electron microscopy (SEM, Figure 1d). These images confirmed the presence of thin phosphorus flakes with lateral dimensions between 50 nm and 50 μm. From these results, we concluded that a more extensive study was required to identify conditions that maximized the yield of thin phosphorus flakes.

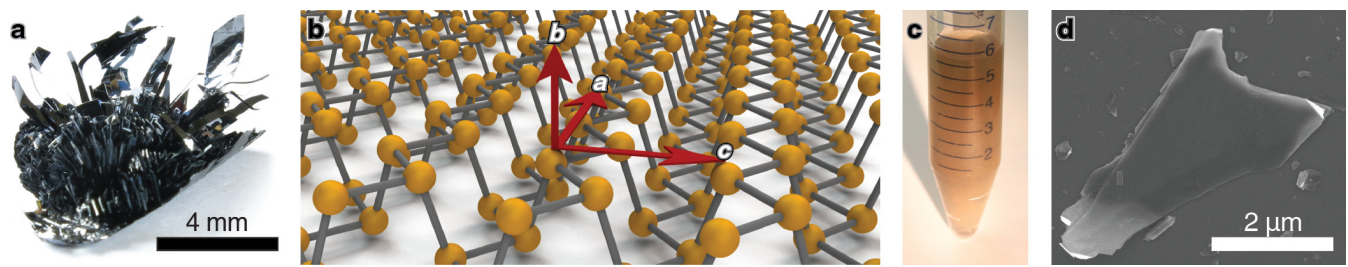

Figure 1. Liquid exfoliation of black phosphorus. (a) Photograph of black phosphorus grown by chemical vapor transport. (b) Illustration of a phosphorene monolayer showing the conventional crystallographic axes. The zig-zag direction is 'a', the armchair direction is 'c', and the 'b' direction is normal to the flake. (c) Photograph of a liquid-exfoliated suspension of 2D phosphorus in isopropanol. (d) SEM image of liquid-exfoliated 2D phosphorus.

We surveyed[42] eighteen solvents for their ability to exfoliate black phosphorus (see Supporting Information for full experimental details). Black phosphorus (10 mg) was added to 20 mL of each solvent and sonicated for thirteen hours under anhydrous and air-free conditions. The suspensions were centrifuged at 3,000$g$ for 30 minutes to remove unexfoliated black phosphorus. The supernatant was further purified *via* dialysis to remove small (< 2.5 nm) phosphorus fragments. These suspensions were characterized with inductively coupled plasma-mass spectroscopy (ICP-MS) and UV-vis transmission spectroscopy to measure a dispersed concentration. We found that the best solvent was benzonitrile, which achieved a mean concentration of 0.11 ± 0.02 mg/mL. Plots of phosphorus concentration *vs*. the Hansen solubility parameters of each solvent (Figure 2a-d) allow us to estimate that the Hildebrand parameter for 2D phosphorus is 22 ± 3 MPa$^{1/2}$. Although there is significant solvent-to-solvent variability—a feature common to graphene, boron nitride, and transition metal dichalcogenides[41]—we find that the optimal solvents for 2D phosphorus are similar to those for other 2D materials. An essential difference, however, is that 2D phosphorus must be handled and sonicated under an inert atmosphere, as we demonstrate below.

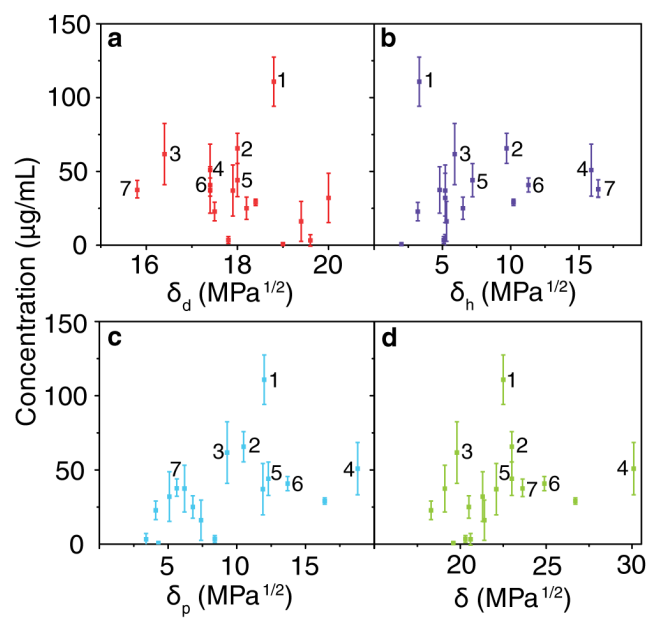

Figure 2. Survey of organic liquids, showing 2D phosphorus concentrations *vs*. Hansen (a-c) and Hildebrand (d) solubility parameters for 18 solvents. The Hansen plots depict the energy due to dispersion forces (a), hydrogen bonding (b), and dipolar intermolecular forces (c). Numbers 1 through 7 rank the best liquids: (1) benzonitrile, (2) 1,3-dimethyl-2-imidazolidinone, (3) 1-vinyl-2-pyrrolidinone, (4) *N*-methylformamide, (5) *N*-methyl-2-pyrrolidone, (6) *N,N*-dimethylformamide, (7) 2-propanol. Each data point is an average of three trials; the error bars correspond to the standard deviation.

*Characterization of 2D phosphorus*

In order to examine the structure of the suspended material, we used transmission electron microscopy (TEM) to quantify shape, size, and thickness as well as high resolution TEM (HR-TEM) to assess crystallinity. We imaged and measured thousands of phosphorus flakes; Figure 3a-c shows TEM images of several representative samples. As before, a broad distribution of flake sizes was found. Single pieces typically had uniform contrast, suggesting that they had a planar morphology. All of the pieces examined in HR-TEM exhibited lattice fringes, showing that the crystallinity of phosphorus flakes was preserved (Figure 3d). We analyzed HR-TEM images by performing fast Fourier transforms (FFT), allowing us to observe the expected {200} and {002} plane families of black phosphorus. In addition, some flakes exhibited strong 101 intensities (Figure 3e), which are forbidden sets of diffracting planes in bulk black phosphorus. To understand the origin of the 101 spots, we used multi-slice calculations (JEMS[47]) to simulate HR-TEM images of 2D phosphorus sheets with varying thicknesses from four common microscopes (see Supporting Information for additional details). Fast Fourier transforms were applied to the HR-TEM images to determine the intensities of spots corresponding to plane families. In agreement with a previous analysis of electron diffraction patterns,[28] we found that a large 101:200 intensity ratio in FFTs is a unique characteristic of monolayers (Figure 3f) when imaged at or near Scherzer defocus, thus confirming their presence in our suspensions. We attribute the diffuse background of the FFT (Figure 3e) to the likely presence of absorbed organics, which has been observed previously for other 2D materials that were not degassed at elevated temperatures prior to imaging.[48,49]

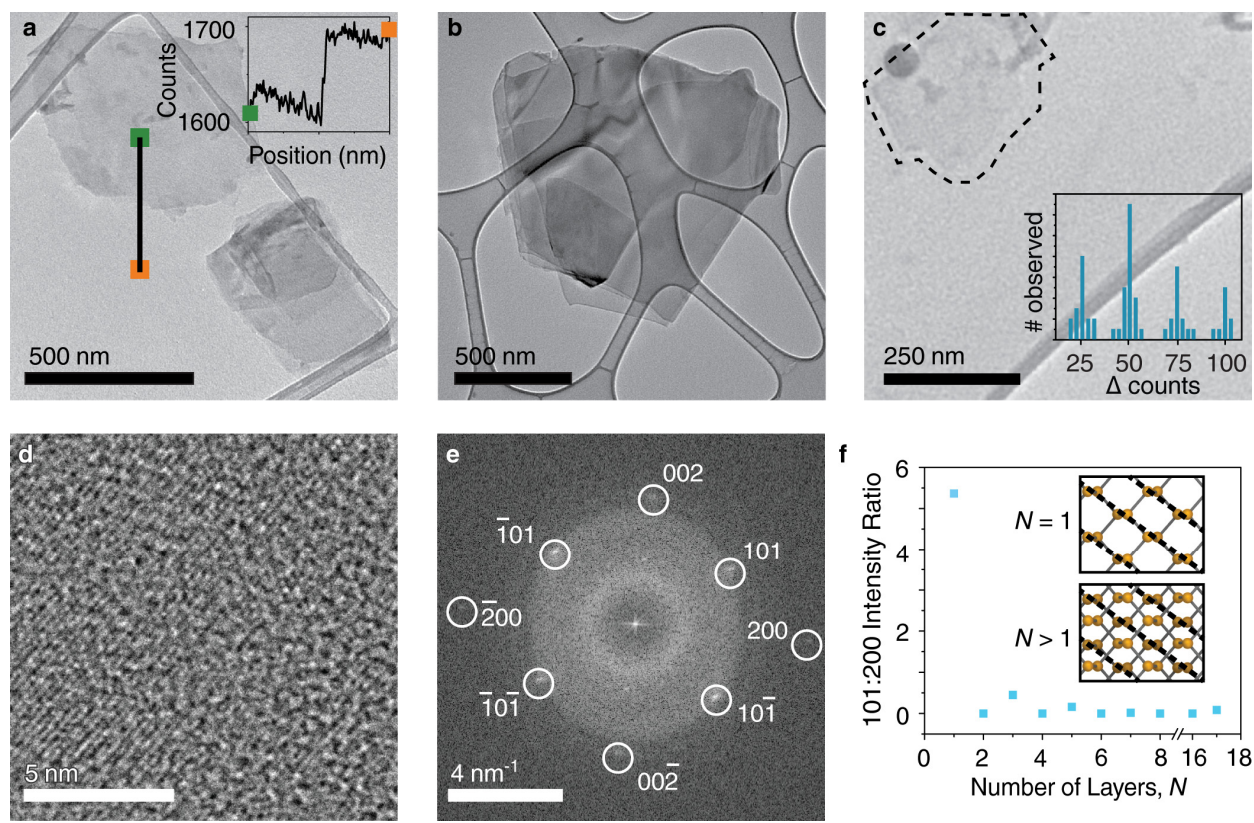

Figure 3. TEM characterization of liquid-exfoliated 2D phosphorus. (a-c) TEM images of 2D phosphorus. (c) TEM image of a monolayer of 2D phosphorus. The inset in (a) shows the contrast change (*ca.* 75 counts) from a line profile drawn across a flake that is three layers thick. The inset in (c) provides a histogram of contrast changes from one hundred flakes. The changes in intensity (25, 50, *etc.*) correspond to monolayers, bilayers, *etc.* (d) HR-TEM image of phosphorene, a monolayer. (e) FFT of the HR-TEM image in (d). (f) Intensity ratios of 101 and 200 spots in FFT HR-TEM images and their relation to layer thickness, as calculated from multi-slice simulations in JEMS.[47] Insets show that constructive interference from {101} plane families (dashed lines) occur in monolayers but have low or no intensity in multilayer flakes.

To quantify the thickness of all flakes in our suspensions, we used our real-space TEM images—all acquired under identical imaging conditions including exposure time, aperture selection, lens currents, magnification and defocus value—to measure the change in intensity across sheet edges for hundreds of flakes (Fig 3a, line and inset). Flake edges were suspended over either vacuum or carbon film (see Supporting Information for additional details). The smallest intensity change was 25 ± 3 counts and all other intensity changes were multiples of 25 counts (Figure 3c, inset). We therefore assigned an intensity change of 25, 50, 75, and 100 counts to monolayers, bilayers, trilayers, and four-layered 2D phosphorus flakes, respectively. Further confirming this assignment, we found that only those flakes with a contrast change of *ca.* 25 counts had the intense 101 spots that are a hallmark of monolayers. Although this method is simple and fast, we do note that the linear relationship breaks down for flakes that are thicker than *ca.* 40 layers.

With the goal of isolating 2D flakes with well-defined thicknesses and optical properties, we used centrifugation to fractionate the phosphorus suspensions. We centrifuged at a rotational centrifugal force (RCF) as low as 120*g* and then centrifuged the supernatant at a slightly greater RCF, reaching values of up to 48,000*g*. The sediment from the second centrifugation was collected and re-dispersed in pure solvent. This new suspension is labeled by the average centrifugal force between the two RCFs; for example, a suspension labeled 20,200*g* has been centrifuged at 17,200*g* and 23,400*g* (see Supporting Information for full experimental details). Using TEM, we analyzed the thicknesses (Figure 4a) lateral size (Figure 4b) and of the suspended 2D phosphorus flakes. We found that this centrifugation approach could systematically isolate flakes with varying size and thickness distributions. When centrifuging at high speeds, for example, we collected macroscopic quantities of flakes with size distributions centered near one-layer and two-layer thicknesses (Figure 4a) in which monolayers comprised up to 45% of the sample. Phosphorene—a material that has been sought after but rarely observed—is now easily accessible.

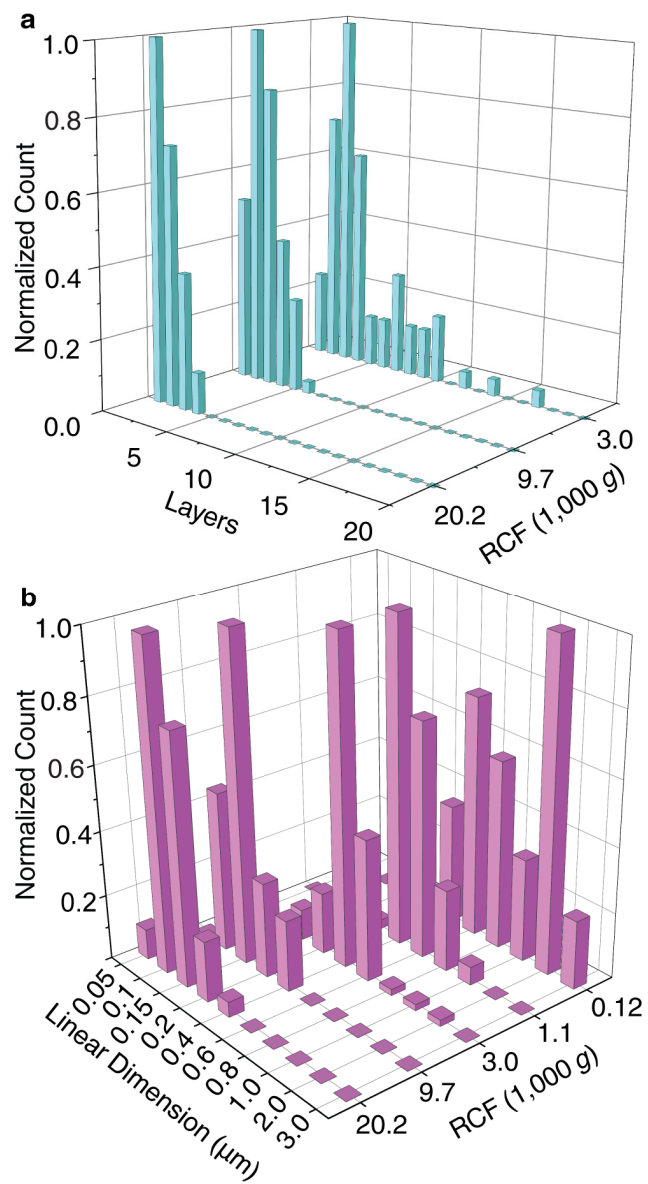

Figure 4. Selective variation of the centrifugation rate allows for control over flake thickness (a) and flake lateral size (b).

*Stability of phosphorene and 2D phosphorus*

As first recognized by Bridgman in 1914,[ref. 14] black phosphorus oxidizes and converts to phosphoric acid under humid atmospheric conditions. More recent studies have also shown that mechanically-exfoliated phosphorus degrades in air.[19,28,32] We used x-ray photoelectron spectroscopy (XPS) and measurements of apparent pH to assess the oxidation. We performed XPS both on bulk black phosphorus to obtain a reference spectrum (Figure 5a) and on 2D phosphorus to test whether oxidation accompanies liquid exfoliation (Figure 5b). In addition to performing all exfoliation and centrifugation under an inert atmosphere, we constructed a transfer chamber that excluded oxygen and water during sample transfer to and from the XPS instrument (see SI for experimental details). Pristine black phosphorus had $2p_{1/2}$ and $2p_{3/2}$ peaks that are characteristic of unoxidized elemental phosphorus.[50] We exposed the same sample to air and re-acquired XPS spectra at later time intervals. A broad peak at 134 eV emerged, which can be attributed to several types of phosphorus-oxygen bonds.[51] Lacking an oxidation mechanism, we cannot yet identify the type or types of P-O species that may be present in our samples. We performed similar experiments on thin 2D phosphorus (< 6 layers). The pristine sample exhibited no signs of oxidation (Figure 5b, black). Upon exposure to oxygen gas that contained some water (not dried) and 460 nm light, a broad peak appeared at 133 eV, characteristic of oxidized phosphorus. In this modified material, *ca.* 5% of the phosphorus was oxidized, as estimated by peak integration software. Collectively, our analyses demonstrate that liquid exfoliation successfully yields high-quality, unoxidized 2D phosphorus.

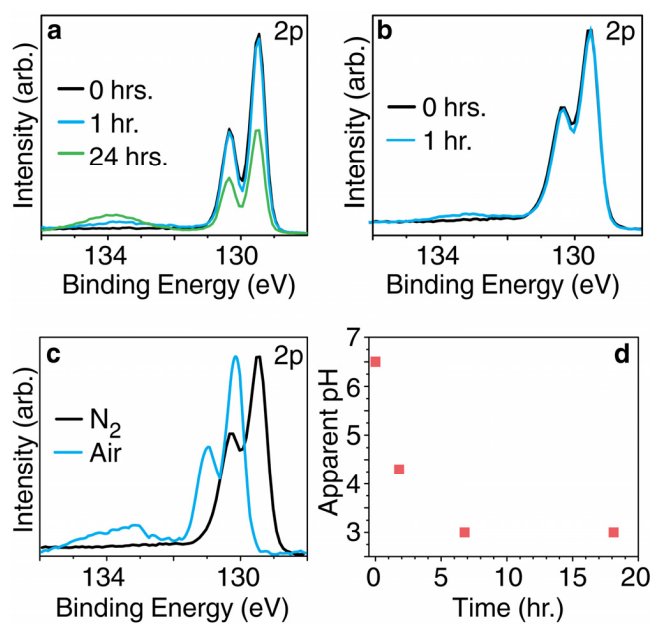

Figure 5. (a) XPS of freshly cleaved bulk black phosphorus after exposure to ambient air and room light for 0 (black), 1 (blue), and 24 (green) hours. (b) XPS analysis of few-layer 2D phosphorus showing that the material prepared by liquid exfoliation was unoxidized (black). The few-layer sheets were controllably oxidized by exposure to light ($\lambda$ = 460 nm) and oxygen with some water (blue). (c) Exfoliation of black phosphorus in a sealed vial with $N_2$ (black) or air (blue) in the head space of the vial shows that the presence of air causes 28% of the phosphorus to become oxidized. Binding energies also increase, although the origin of this effect—whether sample charging, doping, or both—is not yet clear. (d) When few-layer 2D phosphorus (< 6 layers) is suspended in isopropanol and exposed to light and air ($\lambda$ = 460 nm), the apparent pH (recorded by a pH meter) decreases because of acid production.

In order to evaluate whether handling under an inert atmosphere is important, we sonicated black phosphorus in a sealed vial, with either nitrogen or air in the vial's headspace. Analysis of the air-exposed material by XPS (Figure 5c) shows substantial oxidation, with 28% of the phosphorus no longer in the unoxidized form. In addition, we monitored the pH of a solution of few-layer phosphorus that was suspended in isopropanol and exposed to light and air (Figure 5d). We found that the solution rapidly acidifies, consistent with Bridgman's prediction[14] that phosphoric acid is produced upon exposure to air. When higher phosphorus concentrations are used, the apparent change in pH is larger. On the basis of these and prior findings,[43] we conclude that although liquid exfoliation in the presence of air may produce some crystalline, thin material, its surfaces and interior[52] are oxidized, acid is present, and its overall quality is low.

***Exfoliation of black phosphorus at the 10-gram scale***

We explored shear mixing[42,53] as a method for the scaled-up production of 2D phosphorus. We used a Silverson L5M-A shear mixer with either a 0.75-inch or 1.385-inch rotor with square holes for our work at the 1-gram and 10-gram scales, respectively. All experiments were performed under oxygen-free and water-free conditions by bubbling nitrogen gas into the mixing container. In addition, we used a water bath to keep the solutions at room temperature during mixing. We used several different grades of NMP, as it was disclosed to us by the Coleman group that only certain types of NMP may work for shear mixing of graphene.[54] Ultimately, we selected NMP from Sigma Aldrich (99.5% purity, anhydrous) for our scaled-up exfoliation. Black phosphorus was ground in a mortar and pestle prior to its use in shear mixing. We used two different grades of black phosphorus, both of which we produced in our laboratory. The first, "high quality" black phosphorus, was highly crystalline with millimeter-sized crystals and was difficult to grind; the second, "low quality" black phosphorus, was highly polycrystalline, had trace amounts of red phosphorus, and was easy to grind. In our experiments, we found that only the low-quality material could be successfully exfoliated by shear mixing alone, regardless of the type of NMP or the conditions of shear mixing. This observation is consistent with a mechanism in which the separation of layers is nucleated at grain boundaries or other defects in the material. In order to exfoliate the higher-quality starting material, we had to rely on a combination of shear mixing and bath sonication.

For our scaled-up synthesis, we dispersed 6 grams of pulverized, high-quality black phosphorus into 100 mL of NMP and bath sonicated the suspension for 2 hours. Next, we added 700 mL of NMP and shear mixed the sample at 5,000 rpm for 4 hours. The dispersion was sonicated again for 3 hours and then shear mixed again for 1 hour at 5,000 rpm. The resulting suspension is shown in Figure 6a. The material was then centrifuged at 20,200$g$ to yield a highly concentrated suspension of very thin, fractionated material (Figure 6a, small vial). In this suspension, nearly 25% of the sample was monolayers (Figure 6b) and the lateral size (Figure 6c) was similar to the material produced using bath sonication at a smaller scale (Figure 4). This demonstration reveals that the production of high quality 2D phosphorus—including phosphorene—can be readily accomplished using simple and scalable approaches.

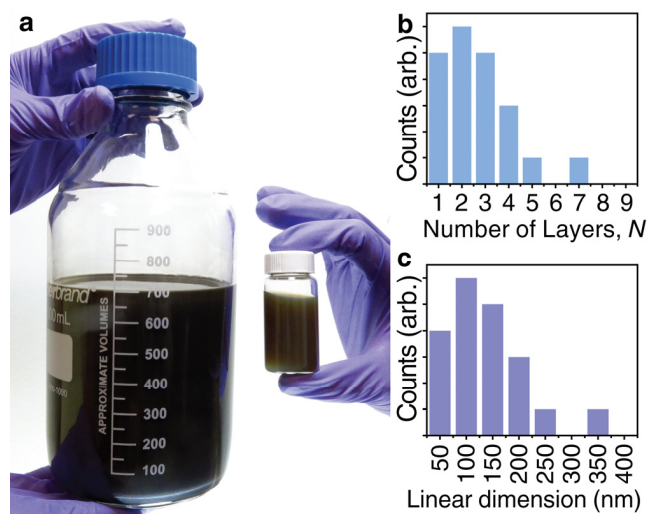

Figure 6. Scaled-up production of 2D phosphorus. (a) Photograph of solutions that were exfoliated using a combination of shear mixing and sonication. In our scale-up, we used six grams of black phosphorus and 800 mL of NMP (left). We centrifuged 40 mL of this mixture at 20,200*g* to isolate a highly concentrated suspension containing thin pieces (right). The size distribution of 2D phosphorus in this fraction is shown in (b) and (c).

*Optical absorption in 2D phosphorus: background*

The optoelectronic properties of black phosphorus and 2D phosphorus—high mobility, anisotropy and the extreme variation in band structure with flake thickness—have provoked intense interest and debate. In 1981, a calculation first proposed the idea of a monolayer of black phosphorus[55] (*i.e.*, phosphorene) and calculated a band gap of 1.8 eV, which is significantly larger than the bulk value of 0.33 ± 0.02 eV (see below for discussion). This remarkable prediction was dormant until several months ago, when the possibility of making phosphorene began to emerge. Despite this interest, the synthesis of monolayers has remained a challenge and, consequently, the majority of recent studies have been theoretical. These studies have essentially confirmed the 1981 prediction—that the band gap is tunable—although they have also introduced considerable uncertainty as to the actual size of the gap: values for monolayers typically range from 1.0 to 2.2 eV (see Table 1). Nevertheless, theory consistently predicts that the band gap is direct for all thicknesses of 2D phosphorus, which has driven further interest because most other 2D semiconductors have indirect band gaps.

These predictions are compelling and need to be systematically examined but, so far, only a few experiments have been reported. Photoluminescence measurements have shown that these predictions are qualitatively correct, but with a varying exciton binding energy of 0.01 to 0.9 eV in phosphorene, this technique will underestimate the band gap of black phosphorus by a similar amount, which depends on the static dielectric constant of the surrounding medium.[56–58] In addition, surface defects, contamination, and oxidation of samples may introduce further experimental uncertainty. In fact, results so far are quite varied: in one study, a trilayer photoluminesced at 1.60 eV, while, in another, the measured value was 0.97 eV (see optical gaps, Table 1). Electrical measurements have also been performed and the reported mobility gaps were smaller than those found by photoluminescence (see mobility and optical gaps, Table 1), a result that is surprising because the mobility gap should be larger than the optical gap in a semiconductor with few interband states.[59,60] However, the study did provide a detailed analysis of many flake thicknesses, and revealed that bulk properties begin to transition towards quantum-confined properties at flake sizes as large as 30 layers.

| Table 1. Reported optical, mobility, and band gaps of 2D phosphorus | | | | | | |
|---|---|---|---|---|---|---|
| **Thickness (layers)** | 1 | 2 | 3 | 4 | Bulk | Source |
| **Photoluminescence (optical gap, eV)** | 1.75 | 1.29 | 0.97 | 0.84 | | Yang, J.[61] |
| | 1.45 | | | | | Liu, H.[16] |
| | 1.31 | | | | | Wang, X.[62] |
| | | 1.29 | 0.98 | 0.88 | | Zhang, S.[63] |
| | | | 1.60 | | | Castellanos-Gomez, A.[28] |
| **Electrical (mobility gap, eV)** | 0.98 | 0.71 | 0.61 | 0.56 | 0.30 | Das, S.[64] |
| **Computation (band gap, eV)** | 2.15 | 1.70 | 1.48 | 1.36 | 1.08 | Castellanos-Gomez, A.[28] |
| | 2.0 | 1.30 | 1.06 | | 0.30 | Tran, V.[37] |
| | 1.94 | 1.7 | 1.3 | 0.8 | 0.43 | Liang, L.[65] |
| | 1.60 | 1.01 | 0.68 | 0.46 | 0.10 | Rudenko, A.[66] |
| | 1.52 | 1.01 | 0.79 | 0.67 | 0.36 | Qiao, J.[67] |
| | 1.01 | 0.66 | 0.52 | 0.47 | 0.31 | Liu, H.[16] |
| | | 1.02 | 0.79 | 0.68 | | Zhang, S.[63] |
| **Absorbance (band gap, eV)** | | 1.88 | 1.43 | 1.19 | 0.33 | This work, see Table 2. |

In this section, we report our experiments on the optical absorbance of black phosphorus and fractionated suspensions of 2D phosphorus. We also report our analyses of these spectra, from which we estimate the absorption edge and band gap in black phosphorus and 2D phosphorus. Some of our analysis uses Elliot's theory of light absorption[68] by delocalized, Wannier-type excitons,[69] and we implement Elliot's theory in the form of Tauc plots.[70] Tauc plots determine the band-to-band transition energy as well as the nature of the transition—whether it is phonon-mediated (indirect) or not (direct), and whether it is dipole-mediated (allowed) or not (forbidden). A proper Tauc plot yields a linear relationship between $(\alpha h\nu)^n$ and h$\nu$, where $\alpha$ is the absorption coefficient, h$\nu$ is the photon energy, and *n* describes the nature of the transition. Although Tauc plots have been criticized because of their simplistic assumptions about band structure and their poor treatment of excitonic effects,[56] they have been used to analyze the absorption edge of many semiconductors, including black phosphorus.[71] In a reported Tauc analysis of black phosphorus, a room-temperature band gap of 0.31 eV was found.[71] This agrees with previously reported electrical measurements,[17,72–76] which we have averaged to calculate a room-temperature band gap of 0.33 ± 0.02 eV. Although this agreement is promising, there are important differences between our 2D samples and bulk black phosphorus that may prevent the application of Tauc's method to our materials. Next, we consider these differences and the corresponding limitations of Elliot's theory.

We have identified five possible reasons why a Tauc analysis could fail to apply to our 2D phosphorus suspensions.

(1) *Light scattering:* A Tauc analysis requires an accurate measurement of the absorption coefficient, α, *versus* wavelength. We measured light that is absorbed by our suspensions of 2D phosphorus using a transmission geometry, but in a traditional transmission geometry, most scattered light is not captured by the detector. To account for forward-scattered light, we placed samples near the opening aperture of an integrating sphere. This measurement showed that the amount of forward-scattered light was relatively small. In addition, because there is less back-scattered light than forward-scattered light,[77] we estimated that our measurements that capture both the transmitted and forward-scattered but neglect back-scattered light have less than a 3% error (see Supporting Information for complete details). Consequently, we have reported an absorption coefficient rather than an extinction coefficient.

(2) *Exciton binding energy:* Elliot's theory is only applicable to Wannier excitons, which have an exciton binding energy (EBE) of less than 100 meV. Bulk black phosphorus has an EBE of 8 meV and the excitonic features in absorbance spectra are only apparent at low temperature.[17] The predicted EBE of phosphorene (a monolayer) depends on the static dielectric constant of the surrounding medium, and can be as large as 900 meV in a vacuum.[28,37] We performed most optical absorbance experiments in NMP, which has a high dielectric constant (32.17) and yields a small EBE (15 meV, see Supporting Information). The small EBEs, combined with the measurement of our absorbance spectra at room temperature and low light intensities, allows Elliot's theory to be applied because excitons will not obscure the absorption edge as they do in $MoS_2$ and other transition metal chalcogenides.

(3) *Urbach tail:* In materials with significant structural disorder, a pronounced absorption extends below the absorption edge.[78] This absorption, called an Urbach tail, could be present in 2D phosphorus because of the loss of periodicity and presence of defects at the edge of sheets. Urbach tails give a non-linear contribution to Tauc plots. To avoid misinterpreting our spectra, we only extracted an estimate of the band gap when a linear fit of the Tauc plot was obtained at energies above the Urbach tail.

(4) *Anisotropic optical properties:* The nature of black phosphorus' band gap depends on direction: it is direct and allowed in the *c* direction but direct and forbidden in the *a* direction (see Figure 1b).[17] In principle, this would prevent a Tauc plot from distinguishing either transition. Fortunately, the forbidden transition is relatively weak and its contribution to light absorption is negligible;[17] thus, it does not obscure the Tauc analysis of the direct, allowed band gap.

(5) *Variation in band gap:* Tauc analyses are typically applied to materials with a single band gap. If multiple gaps are present and they span a narrow range of energies, it is not possible to distinguish each gap. The superposition of multiple absorption edges of similar strength leads to non-linearity in the Tauc plot, preventing one from determining the nature of the absorption edge or from extracting an accurate band gap energy. Of the five limitations, we found that this consideration is the most important. Our suspensions contain flakes of several thicknesses and therefore several band gaps. Because the absorption coefficients from flakes of different thicknesses are similar and because their band gaps fall across a range of energies, we found that it is not always possible to use a Tauc analysis (see below). In those instances, we have developed and applied a different method for estimating the absorption edge.

***Optical absorption in 2D phosphorus: measurement and Tauc analysis***

In this section, we report our measurement and Tauc analyses of the optical absorbance of 2D phosphorus suspensions. In order to interpret these measurements, we first established reference spectra of bulk black phosphorus. We performed UV-vis-nIR (175 nm to 3,300 nm) and FT-IR measurements on a polycrystalline sample (KBr pellet, Figure 7a, black) and used a CRAIC microspectrophotometer on single flakes of mechanically cleaved bulk crystals (Figure 7a, gray). All spectra were acquired under an inert atmosphere. Fractionated suspensions of 2D phosphorus (Figure 4) were analyzed using an integrating sphere to capture both transmitted and scattered light (see Supporting Information for full experimental details).

The polycrystalline black phosphorus within the KBr pellet had a high optical density, which allowed us to quantify light absorption near the band gap threshold. We observed an onset of absorption at *ca.* 0.4 eV (Figure 7a, black), characteristic of bulk black phosphorus. The analysis of cleaved phosphorus flakes (20 to 40 nm thick) with low optical density revealed an additional absorption edge at *ca*. 1.95 eV (Figure 7a, gray). We attribute this absorption event to a higher energy transition. We will see that these two absorption thresholds—the low-energy band gap transition and the high-energy transition (see band diagram in Figure 8b)—are also present in suspensions of 2D phosphorus.

The fractionated suspensions of 2D phosphorus varied significantly in their appearance: in transmitted light, dilute suspensions of thick pieces appeared black or brown while those containing primarily thin pieces appeared red or yellow (Figure 7b, inset). These observations were consistent with the corresponding optical absorbance spectra of the suspensions (Figure 7b) in which we observed a spectral blue-shift as the flake thickness decreased. There are two notable features in these spectra: a sharply rising absorption within the visible region and a slowly rising absorption that extends into the near-IR. In the following analysis, we will attribute these spectral features to the same high- and low-energy transitions observed in the bulk material.

We sought to quantify these absorption features by using a Tauc analysis. The high-energy transition achieved an excellent fit to a Tauc model when $n = 2$, indicating that this transition is direct and allowed (Figure 7c). We assigned the high-energy transition energies to values of 1.95 eV in bulk black phosphorus and 3.15 eV in a suspension containing primarily monolayers, the thinnest fraction analyzed. The fact that the Tauc models fit our data may suggest that the five experimental challenges outlined above—light scattering, high exciton binding energy,

Urbach tail, anisotropic optical properties, and variation in band gap—have a negligible effect on our Tauc analyses of the high-energy transition.

When we applied the direct Tauc model to the low-energy transition, we measured a value of 0.40 eV for bulk black phosphorus (Figure 7d, bulk) which is slightly larger than previous estimates of its band gap. The method of sample preparation—grinding bulk black phosphorus with KBr to make a pellet—may have exfoliated some thin sheets, yielding a slightly larger band gap. We found that the band gap is direct and allowed, which is consistent with earlier findings. Because theory consistently predicts that the band gap is direct for all thicknesses of 2D phosphorus, we attempted to apply direct Tauc models (both allowed and forbidden) to the low-energy, band gap transition of 2D phosphorus. For all Tauc models that we explored, we never found a linear region of the Tauc plot, which prevented us from determining the band gap using this method (Figure 7d shows the direct, allowed Tauc plot). We attribute the non-linearity of the Tauc plot to several causes. First, the low-energy transition has a lower absorption coefficient than the high-energy transition. The weak absorbance is more likely to be obscured by other optical processes, such as light absorption from Urbach tails or light scattering. Second, the polydispersity of our samples gives a broader distribution of absorption edges for the low-energy absorption than the high-energy absorption. This is because the high-energy transition is less sensitive to flake thickness than the low energy transition, as will become apparent in the following analysis. Because of these experimental challenges in applying a Tauc analysis to the absorption edge of 2D phosphorus, we introduce a new analytical method that can supplant the Tauc method when analyzing families of bulk and quantum-confined semiconductors.

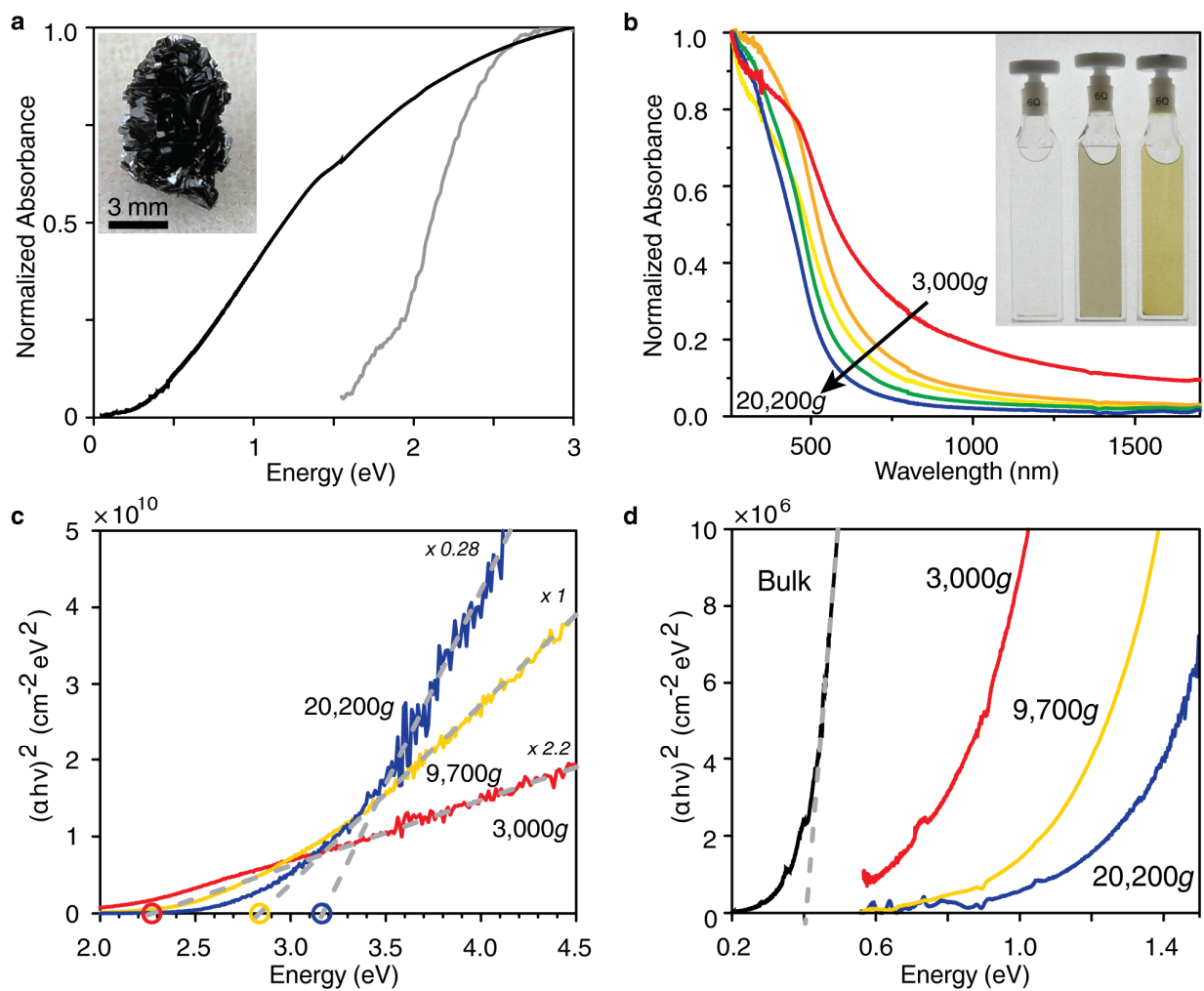

Figure 7. UV-vis-nIR spectroscopy of black phosphorus and its liquid-exfoliated few-layer flakes. (a) Optical absorbance of bulk phosphorus measured at two different optical densities (black = high; gray = low) to reveal two distinct optical transitions (*ca.* 0.4 eV and 1.95 eV). (b) Absorbance of 2D phosphorus suspensions that were prepared by fractionation at RCFs near 3.0, 5.9, 9.7, 14.5, and 20.2 thousand *g*'s (red to blue). (c) Representative direct Tauc plots used to determine the band-to-band transition. (d) Representative direct Tauc plots of the low-energy optical transition. The fit to Tauc models is poor, consistent with the wider range of optical absorption edges that are present in these suspensions.

*A method for determining absorption edges in quantum-confined semiconductors*

In our suspensions of 2D phosphorus, sample polydispersity has prevented a straightforward application of Elliot's theory. Indeed, this is an extremely common problem and the liberal application of the Tauc method often causes large errors in the measurement of band gaps.[56] To circumvent these challenges, we now introduce an alternate method that can be applied to families of bulk and quantum-confined semiconductors such as black and 2D phosphorus. We validated our method using simulated absorption spectra of monodisperse and polydisperse suspensions of 2D phosphorus. Our tests demonstrate that the method is robust: it can determine absorption edges of semiconductors in polydisperse samples and has several advantages over the Tauc method, such as providing an estimate of uncertainty in the absorption edge energy (usually less than a few percent). Crucially, the measurement of an absorption edge (also called the optical gap) also allows us to determine the band gap because the optical and band gaps differ in energy by the exciton binding energy, which is <15 meV in our experiments and therefore negligible.

Our analytical method, which we call the "alpha method", utilizes the similarities that often exist between the electronic structures of quantum-confined semiconductors and the corresponding bulk semiconductor. As an example, numerous studies of black phosphorus and 2D phosphorus show that the band gaps of bulk and 2D phosphorus are always direct with both allowed and forbidden contributions, that their lowest energy transition is always located at the Z-point (in a 3D Brillouin zone), and that the conduction and valence bands are always comprised primarily of $p_z$ orbitals.[17,37,55] In the case of the black phosphorus family, these similarities result in joint densities of states near the absorption edge that are virtually unchanged among members of the family, except for an effective scissoring of the band gap energy. In general, the absorption coefficient increases with increasing quantum confinement,[79] but we hypothesized that the change in the absorption coefficient at the absorption edge ($\alpha_{AE}$) with confinement would be small and could therefore be treated as being unchanged from the bulk to the monolayer. While this is an oversimplification, we will show that this introduces only a small error in the determination of optical/band gap energies.

The step-wise analytical method that follows from this hypothesis is illustrated in Figure 8a. Using black phosphorus as an example, we exploit the fact that the band gap of the bulk material has been measured many times and has a well-defined value (0.33 ± 0.02 eV). First, we measure the absorption spectrum of the bulk material to determine the value of $\alpha_{AE}$. Second, we measure the absorption spectra of a series of samples of 2D phosphorus. Finally, we assign the band gap of each sample as the energy at which the absorption coefficient equals $\alpha_{AE}$ (see Figure 8a). Note that this process is equating the absorption edge (optical gap) and the fundamental absorption edge (band gap), which is an accurate approximation in our experiments but is not necessarily true in all cases.

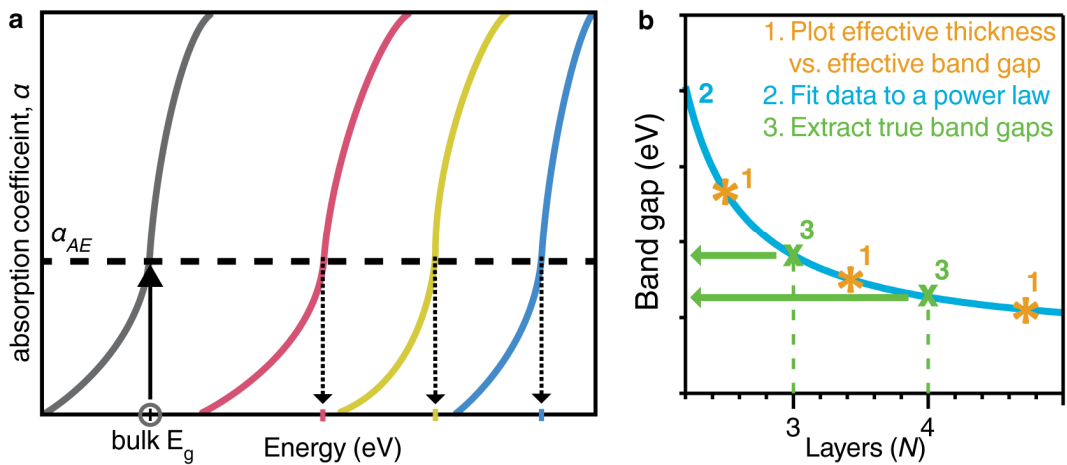

Figure 8. "Alpha method" for band gap determination. (a) The absorption coefficient at the absorption edge ($\alpha_{AE}$) is measured for the bulk material. We use this $\alpha_{AE}$ to estimate the band gap energy for the quantum-confined 2D flakes. If the 2D flakes are not monodisperse in thickness, the band gap that we have determined is an effective band gap. (b) To convert an effective band gap into a real band gap, we (1) plot the effective band gap *vs*. effective thickness for a series of polydisperse samples. Next (2), we fit the data to a power law, see equation [1] in the text. Lastly (3), we use the power law fit to extract the band gap for 2D flakes with real thicknesses.

To validate the alpha method, we used four calculated ($G_0W_0$) absorption spectra—the spectra that come from bulk, trilayer, bilayer, and monolayer phosphorus.[37] For each spectrum, we used the reported band gaps, acquired by measuring the energy difference between the conduction and valence bands. We then applied the alpha method to the same data to obtain a second estimate. Across this family of materials, we found that the maximum difference between the methods was 1.85%—an amount that is essentially negligible for most purposes (see Supporting Information for a complete analysis). Although the central assumption—that $\alpha_{AE}$ is the same in all members of the family—is not true, the error due to this assumption is small. This is because $\alpha$ rises steeply near the band gap ($d\alpha/dE$ is large as the energy $E$ approaches the band gap energy $E_g$). Consequently, even if large differences in $\alpha_{AE}$ exist among the members of a semiconductor family, these produce small differences in the estimated band gap energy.

With this set of results for monodisperse samples in hand, we then tested whether the alpha method could be applied to polydisperse samples. We constructed a series of 24 different artificial mixtures of 2-, 3-, 4-, and 5-layer pieces by taking linear combinations of the calculated absorption spectra of the individual flakes. For these mixtures, Tauc plots were often unusable: the plots either contained several linear regions or did not contain any linear region at all. On the other hand, when we applied the alpha method to each suspension, we always obtained an estimate of an "effective band gap". We found that the effective band gap increased monotonically as the sample distributions shifted from containing a majority of thicker flakes (4- or 5-layers) to a majority of thinner flakes (2- or 3-layers) and that, as expected, the effective band gap always fell between the band gaps of the thinnest (2-layer) and thickest (5-layer) flakes. This example shows that the effective band gap does not necessarily correspond to the band gap of any real material but rather represents the contributions from various-sized flakes in a given mixture. Nevertheless, if this effective band gap is properly correlated with an "effective thickness" and these band gap-thickness correlations are performed on multiple samples, then it would be possible, at least in principle, to interpolate between these data points to obtain the band gap of 2D materials with real thicknesses (*e.g.*, a bilayer or a trilayer) as illustrated in Figure 8b.

The challenge with this approach is that it is not obvious whether the interpolated values are correct. We addressed this challenge by applying a robust mathematical approach to determine an effective thickness for each simulated mixture that, when paired with the effective band gap (alpha method), would lead to correct values of the band gap for real flake thicknesses. We compared our effective thicknesses and effective band gaps to those predicted by a power-law fit of the true thickness and true band gaps for the individual flakes in our simulated mixtures. A power-law fit was selected because, as suggested by numerous calculations, it appears to correctly describe the variation in band gap with flake thickness.[16,28,37,67,80] The power law model yields a band gap for the $N_{th}$ layer as:

$$E_{g_N} = \frac{E_{g_1} - E_{g_\infty}}{N^x} + E_{g_\infty} \tag{1}$$

where $E_{g_1}$ is the band gap of phosphorene (a monolayer), $E_{g_\infty}$ is the band gap of bulk black phosphorus and $x$ is a parameter describing the nature of quantum confinement in the system. Values of $x$ are usually between 0 and 2, where the variation is due in large part to the extent of Coulomb interactions[79,81] and therefore depends on the material geometry (quantum dot *vs*. nanowire *vs*. 2D flake). In the present case, the power law fit is useful because it provides an excellent fit to the calculated $G_0W_0$ spectra and because it allows us to make direct comparisons of the real band gaps to the effective band gaps at non-real (*i.e.*, non-integer) thicknesses.

In these calculations, we considered the same series of mixtures as above. The skewness of these 24 distributions was systematically varied to capture the full range of likely skews that may be observed experimentally, which, as seen in Figure 4, typically have a log-normal shape. For each artificial mixture, we employed the alpha method to determine an effective band gap and we tested five different statistical approaches to extract effective thicknesses. The approaches that we tested were a number-averaged mean (analogous to $\overline{M_n}$ in polymer physics), a weight-averaged mean (analogous to $\overline{M_w}$ in polymer physics), and the mean, median and mode that were derived from a log-normal fit to each distribution. We note that the weight-averaged mean is not equivalent to a weight fraction, which is defined as the weight of material per total weight of solvent and material (see Supporting Information, Section 11, for a complete description of these statistical measures).

From these 24 mixtures comprising realistic skews, we found that the best two averages were the log-normal mean and the number-averaged mean. When paired with the effective thickness as calculated by the log-normal mean, the calculated band gap (power law) was 0.3 ± 1.5% above the effective band gap (alpha method). When paired with the effective thickness from the number-averaged mean, the calculated band gap was 0.2 ± 2.6% below the effective band gap. The next two closest measures of thickness were the log-normal median (2.3 ± 1.6% above the effective band gap) and the weight-averaged mean (3.6 ± 3.5% below the effective band gap). In general, we found that the extent to which these statistical measures over- or underestimated the true bandgap varied systematically with the skewness of the distribution. For distributions with low skewness, the band gap was systematically overestimated by about 1.5%, while distributions with high skewness, such as those obtained in our experiments (Figure 4), the band gap was systematically underestimated by about 1%, although there were a small number of outliers with errors up to 6%. A table and graphs that summarize these calculations are provided in section 11 of the Supporting Information.

The central conclusion from these simulations is that the alpha method, when combined with an appropriate flake thickness, yields band gaps that are reliable. As noted above, the maximum difference between the reported band gap and the alpha band gap was 1.85%. In addition, the maximum error in using either the number-averaged mean or the log-normal mean was 6%. We emphasize that these are maximum errors and the typical errors will be less. However, these estimates of error only describe those errors due to data analysis and do not include systematic or non-systematic errors that are inherent to the experimental measurements.

*Thickness-dependent absorption edges of black and 2D phosphorus*

In this section, we compile the results of our experimental determination of the absorption edge of 2D phosphorus. As we described above, the absorption edge probed by our experiments (the optical gap) is indistinguishable from the band gap because the exciton binding energy is extremely small (8 to 15 meV) and exciton fission is rapid. When discussing the energy associated with a particular transition, we will use "absorption edge", "optical gap" and "band gap" interchangeably.

From the absorption spectrum of bulk black phosphorus, we measured $\alpha_{AE}$ to be 0.24 μm$^{-1}$, which is equivalent to a light penetration depth of 4 μm. (We note that the absorption coefficient determined by us is similar to the one reported previously,[73] $α_{AE}$ = 0.17 μm$^{-1}$.) We then used this absorption coefficient to determine the effective band gap of each 2D phosphorus suspension. The thickness distribution of each phosphorus suspension was analyzed by TEM and this distribution was converted into effective thicknesses using the four most accurate statistical averages (log-normal mean, number-averaged mean, log-normal median and weight-averaged mean). For each of these averages, a power law (Equation 1) was fit to the thickness-band gap data set. The same process was repeated for the high-energy transition, with the only difference being that the band-to-band transition energy was taken from a Tauc analysis rather than the alpha method (see justification, above).

Figure 9 summarizes our most important findings: the experimental quantification of the band-to-band transitions of 2D phosphorus. Figure 9a displays the band gaps (orange curves, "low energy") and high-energy transitions (blue curves, "high energy"). The low energy and high energy transitions show the most probable values (dark orange, dark blue) and a maximum likely range of values (light orange, light blue). The four curves that define the most probable and maximum likely boundaries come from the power-law fits to the four types of effective thicknesses, with the most probable boundaries defined by the log-normal and number-averaged means and with the maximum likely boundaries defined by the log-normal median and weight-averaged mean. The average exponent $x$ of the power law (Equation 1) calculated from our data is 0.81. Earlier theoretical predictions suggest that the value may between 0.7 and 1.0, with an average value of 0.80 ± 0.11 reported across six studies.[16,28,37,66,82,83]

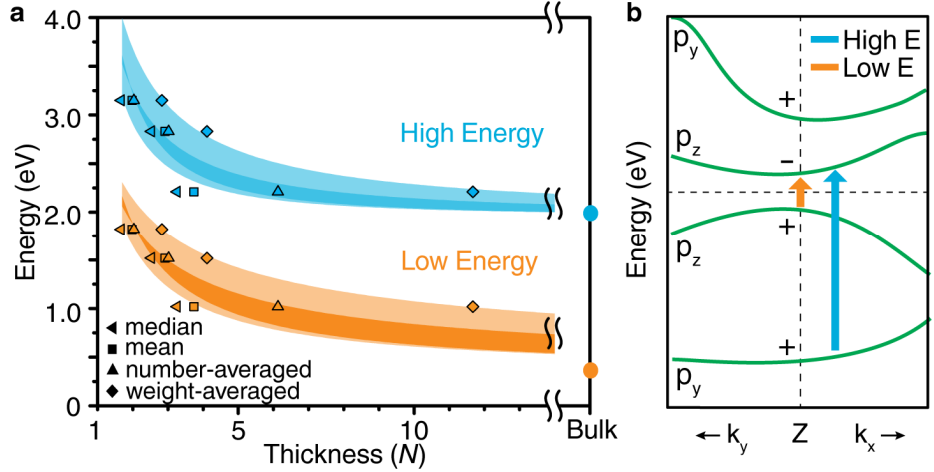

Figure 9. Experimentally determined band gap (low energy) and high-energy transitions of 2D and bulk black phosphorus. (a) The band gap (orange, "low energy") and high-energy band-to-band (blue, "high energy") transitions are plotted with respect to flake thickness. The dark blue and dark orange regions define the most probable energy values and the light orange and light blue define the maximum likely range. (b) Band structure of bulk black phosphorus at the Z point of the first Brillouin zone. The orange arrow represents the band gap transition (VB → CB) while the blue arrow represents the high energy transition (VB-1 → CB). The plot also shows the parity of bands near the Z point (+, -) and the nature of orbitals that primarily contribute to each band ($p_y$, $p_z$). The valence bands come from angle-resolved photoelectron spectroscopy measurements[18,85] and the conduction bands come from calculations.[18,84] The horizontal dashed line shows the Fermi level.

Our measurements provide direct experimental evidence that the band gap and the high-energy transitions undergo extreme changes as flakes approach monolayer thickness. The band gap can be tuned from 0.33 ± 0.02 eV in bulk to 1.88 ± 0.24 eV in bilayers. The higher energy transition can be tuned from 1.95 ± 0.06 eV in bulk to 3.23 ± 0.39 eV in bilayers. These ranges surpass all known 2D materials and are as large as the most tunable quantum dots. The most important band gaps and high-energy transitions are reported in Figure 9a and Tables 1 and 2. Our estimates of error (*e.g.*, ± 0.24 eV for the band gap of bilayers) are the same as the maximum ranges in Figure 9a (light blue and light orange regions). These estimates of error do not include the 1.85% maximum error between the alpha method and reported gaps or experimental error. We have not extrapolated our power law to a monolayer thickness because of the errors associated with such an extrapolation: the smallest effective thickness of our samples was 1.71 layers and it has also been suggested that a phosphorene monolayer does not lie on the power-law curve.[37]

**Table 2. Electronic band-to-band transitions in 2D phosphorus.**

| Layers | Band gap (eV) | High-energy (eV) |
|---|---|---|
| 2 | 1.88 ± 0.24 | 3.23 ± 0.39 |
| 3 | 1.43 ± 0.28 | 2.68 ± 0.32 |
| 4 | 1.19 ± 0.28 | 2.44 ± 0.27 |
| 5 | 1.04 ± 0.27 | 2.31 ± 0.23 |
| 6 | 0.94 ± 0.26 | 2.23 ± 0.20 |
| 7 | 0.87 ± 0.26 | 2.18 ± 0.17 |
| 8 | 0.81 ± 0.25 | 2.14 ± 0.16 |
| 9 | 0.77 ± 0.24 | 2.11 ± 0.14 |
| 10 | 0.73 ± 0.23 | 2.09 ± 0.13 |
| 15 | 0.62 ± 0.20 | 2.03 ± 0.09 |
| 20 | 0.56 ± 0.18 | 2.01 ± 0.07 |
| ∞ | 0.33 ± 0.02 | 1.95 ± 0.06 |

In order to place these measurements in context, we compare our band gaps to prior optical gap measurements (see Table 1). We focus, in particular, on the optical gaps of Yang[61] and Zhang[63] because these studies surveyed the largest range of flake thicknesses and because these studies are the only two that are in agreement. When the optical gap and band gap are measured in the same dielectric environment, the band gap is expected to be larger than the optical gap by an amount equal to the exciton binding energy. This relationship only holds true for measurements that are performed in media with the same dielectric constant, since the exciton binding energy, optical gap, and band gap all depend on the medium's dielectric constant. This sensitivity to the medium's dielectric constant disappears as flakes become thicker and in the limit of thick flakes, the optical gaps and band gaps converge because the exciton binding energy is 8 meV in bulk black phosphorus.[17] To see whether the electrical gap and optical gap do converge, we focus on four- and five-layer thicknesses, which are the thickest flakes that have been studied in the photoluminescence (optical gap) experiments. The four-layer optical gap was reported as 0.86 eV and the five-layer optical gap was 0.80 eV. We measured a four-layer band gap as 1.19 ± 0.28 eV and a five-layer band gap as 1.04 ± 0.27 eV. It is apparent that the difference between the band gap in our experiments and the optical gap in the photoluminescence experiments is decreasing (0.33 eV for four-layer, 0.24 eV for five-layer), as expected. Although the extent to which we can make comparisons is limited by the available data, it appears that there is reasonable agreement between our measurements and some previous photoluminescence measurements.

Next, we turn our attention to the high-energy band-to-band transition. Although this transition has been neglected in earlier studies, we suggest two reasons that understanding this transition will be important. First, the changes in the color of 2D phosphorus with decreasing thickness (Figure 7b, inset) are due, in large part, to changes in the high-energy band-to-band transition rather than the band gap. As a result, the ability to modulate the material's color requires an understanding of the high energy transition. Second, the high-energy transition has a substantially larger absorption coefficient (3.3 μm$^{-1}$ at 3 eV) and a smaller light penetration depth (300 nm at 3 eV) than the band gap transition. This feature will be important in designing 2D phosphorus for applications that require high light absorption. From our Tauc analyses of the high-energy transition, we found that the bulk material has a transition energy of 1.95 eV, increasing up to 3.23 ± 0.39 eV in bilayers (Figure 9b, blue). From these measurements, it is also apparent that the high-energy transition is less sensitive to flake thickness as compared to the low-energy band gap transition: from bulk to bilayers, the band gap changes by 1.55 eV while the high-energy transition changes by 1.28 eV. This difference in sensitivity may be why Tauc plots appear to work well for the high-energy transition while they do not work for the low-energy transition.

Finally, we describe four key observations that allow us to determine the nature of the high-energy transition. First, the Tauc plots show that there is a linear relationship between $(\alpha h\nu)^2$ and the photon energy (Figure 7c), which is characteristic of a direct, allowed transition. Second, the high-energy transition is sensitive to material thickness, varying from 1.95 eV in bulk to 3.23 eV in bilayers. These changes follow a power law and therefore appear to be driven by quantum confinement.[79,81] As such, it is plausible that the high-energy transition occurs at or near the Z-point of the Brillouin zone, since Z is perpendicular to the plane of flakes. In examining the band structure near the Z-point (Figure 9b), it is clear that there are two likely candidates for a direct optical transition at or near the Z-point: a transition between the valence band (VB) and the second lowest unoccupied band (CB+1) or between the second highest occupied band (VB-1) and the conduction band (CB). Third, the energy of the transition for the bulk material (1.95 eV) can be compared to previous measurements of the band structure of bulk black phosphorus.[15,84] From these comparisons, it is clear that only the VB-1 → CB transition provides the right energy. Fourth, the absorption coefficient of the high-energy optical transition is considerably larger (about ten times larger) than that of the low-energy optical transition. This observation is consistent with an assignment of the optical transition to VB-1 → CB: this transition is direct, allowed in the *c*-direction, and leads to a change in parity (+ → -).[17] These selection rules favor strong optical absorption. Figure 9b summarizes this assignment and also identifies the low-energy transition.

**CONCLUSIONS**

In this work, we have described our method[42] for preparing and isolating large quantities of monolayers, bilayers, and few-layer flakes and we identified benzonitrile as the best solvent of those we surveyed. Although shear mixing provides insufficient force for exfoliating high-quality samples of black phosphorus, it is possible to combine shear mixing and sonication to exfoliate black phosphorus at the 10-gram scale. Using XPS, TEM, and multi-slice TEM simulations, we observed that monolayers, bilayers, and few-layer flakes of 2D phosphorus are crystalline and unoxidized. Our work also demonstrates a rapid and simple TEM-based method for measuring the thickness of 2D phosphorus.

Using a method that we introduced here for quantifying the optical absorbance spectra, we showed that it is possible to measure the optical gap of polydisperse 2D phosphorus samples and to extract an accurate estimate of the material's band gap. Our results may go some ways towards resolving the long-standing question of how the band gap of black phosphorus changes with thickness. We expect that the methodology presented here will be broadly applicable as it provides a robust approach for optical or band gap measurement in mixtures of complex semiconductors and can extract useful information even when the Tauc analysis fails.

Of central importance for future applications of 2D phosphorus, we have performed the first accurate measurements of the thickness-dependent band gap. Although there are a large number of theoretical predictions, these predictions have not yet been tested, until now, by careful experiments. We found that the band gap can be tuned from 0.33 ± 0.02 eV in bulk black phosphorus to 1.88 ± 0.24 eV in bilayer phosphorus. It is important to note that the band gap will likely depend on the surrounding medium but, in any case, the range of optical transitions for black and 2D phosphorus is relatively large compared to that of other quantum-confined nanomaterials such as $MoS_2$ (1.2 to 1.9 eV),[9,85] CdSe quantum dots (2.0 to 3.0 eV)[86] or PbSe quantum dots (0.27 to 1.5 eV).[11–13,86] This suggests that the electronic coupling between layers is stronger than in most other van der Waals layered solids but a complete description of this unusual property is still needed. Looking toward future applications of this material, we suggest that the astounding range of band gaps that can be achieved by 2D phosphorus, with tunable absorption thresholds from the infrared to the visible, will provide a new material platform for the design and development of solar cells, photodetectors, photocatalysts, transistors, and batteries.

**METHODS**

*Chemicals*. All chemicals were purchased from Sigma Aldrich unless otherwise noted: black phosphorus (Smart Elements, ~99.998%), red phosphorus (≥ 99.999+%), *N*-methyl-2-pyrrolidone (Sigma Aldrich or Acros Organics, anhydrous, 99.5%) 2-propanol (FisherSci, electronic grade), cyclopentanone (≥ 99%), 1-cyclohexyl-2-pyrrolidone (≥ 99%), 1-dodecyl-2-pyrrolidinone (≥ 99%), benzyl benzoate (≥ 99.0%), 1-octyl-2-pyrrolidone (≥ 98%), 1-vinyl-2-pyrrolidinone (≥ 99%), benzyl ether (≥ 98%), 1,3-dimethyl-2-imidazolidinone (≥ 99%), cyclohexanone (≥ 99.8%), chlorobenzene (≥ 99.5%), dimethylsulfoxide (≥ 99.9%), benzonitrile (anhydrous, ≥ 99%), *N*-methylformamide (≥ 99%), dimethylformamide (≥ 99%), benzaldehyde (≥ 99%). The solvents were degassed prior to use, handled under nitrogen in a glove box ($O_2 < 1$ ppm) and transferred from instrument to instrument using custom-designed glassware with Rotaflo stopcocks. When necessary, instrument loading chambers or entire instruments were enclosed in $N_2$-filled glove bags to provide an inert atmosphere.

*Synthesis*. Black phosphorus was purchased from Smart Elements or made in our laboratory using a method similar to that of Nilges.[46] Red phosphorus (0.5 g), tin (10 mg), and tin iodide (15 mg) were loaded into a partially-sealed quartz tube in a glove box. The tube was 14-cm long, had a 1-cm inner diameter, and a 1.4-cm outer diameter). The tube and its contents were evacuated to *ca.* $4 \times 10^{-3}$ mbar and sealed with an oxy-hydrogen torch while maintaining vacuum within the ampoule. We ensured that the ends of the quartz tubes were >0.4 cm and without entrapped gas bubbles; in one experiment, the quartz tube burst because the walls were too thin. (Caution: only perform this synthesis in a well-ventilated area with restricted access as the risk of unanticipated explosions is high because the tube pressure greatly exceeds 1 atmosphere.) The reaction was carried out in a Lindberg Blue three-zone furnace using the temperature profile described by Nilges. We maintained a 60 °C temperature differential across the 15-cm tube by inserting insulation between the two zones. The tube was opened in our glove box and the reaction products were washed sequentially in anhydrous acetone, *N*-methyl-2-pyrrolidone and isopropanol, and subsequently dried prior to use.

*Liquid exfoliation.* Black phosphorus crystals were ground into granular pieces using a mortar and pestle in a nitrogen glove box. For typical experiments, 10 mg was added to a 20-mL scintillation vial. Twenty milliliters of solvent was added to give a concentration of 0.5 mg/mL. Vials were tightly capped and wrapped with parafilm to prevent air exposure before placing into a Branson 5800 bath sonicator. The bath sonicator was outfitted with a test tube rack to allow for controlled placement of vials. Vials were systematically moved through several locations during the course of sonication to minimize vial-to-vial variations in phosphorus dispersion. The samples were subjected to eight to ten cycles of sonication at high power, each lasting 99 minutes. Bath water was changed after each cycle to maintain a temperature between 22 and 30 °C (during sonication, bath temperature increased dramatically). During the sonication process, the black phosphorus crystals dispersed into the solution and the suspension acquired a brown appearance. After sonication, the vials were returned to the glove box.

*Centrifugation*. Centrifugation was performed in a Sorvall RC-5B superspeed refrigerated centrifuge with a rotor radius of 10.7 cm. Dispersions of sonicated black phosphorus were transferred into Nalgene Oak Ridge FEP 10- or 50-mL centrifuge tubes. We list centrifuge speed as RCF (relative centrifugal force) rather than rpm (revolutions per minute) because rpm alone provides an incomplete description of centrifugation conditions. See Supporting Information for more details on sample handling and procedure.

*Transmission electron microscopy (TEM)*. A JEOL 100CX II TEM was used for low resolution imaging at a 100 kV accelerating voltage with a resolution of 2 Å (lattice) and 3 Å (point to point). For HR-TEM, we used a JEOL 2010F-FasTEM at 200kV accelerating voltage for a resolution of 1 Å (lattice) and 2.3 Å (point to point). Samples were prepared in a glove box, where the phosphorus dispersion (in IPA or NMP) was dropcast (~10 µL) onto a 300-mesh copper TEM grid coated with Lacey carbon (Ted Pella). After deposition, the solvent was evaporated for one hour (IPA) or at least 48 hours (NMP). During this period, TEM grids were held with self-closing tweezers and were not blotted against filter paper to ensure that a representative distribution of the material was deposited. The TEM grids were placed in a grid box wrapped in aluminum foil, placed in a nitrogen-filled zip-loc bag, and brought to the TEM. Grids were rapidly transferred into the TEM with a minimal amount of light exposure. For both TEM analyses, images were acquired and processed with Gatan Digital Micrograph. Fast Fourier Transforms (FFTs) were applied to HR-TEM images to observe the plane families present in the HR-TEM images of phosphorus flakes.

*Edge contrast analysis*. We performed this analysis with the line profilometer tool of Gatan Digital Micrograph software. The lateral dimensions of each flake are reported as the diagonal of two perpendicular length measurements. See Supporting Information for full description.

*Multislice calculations*. To simulate HR-TEM images of 2D phosphorus from one to eight layers, we used Java electron microscopy software (JEMS) version 4.1830U2014 by Pierre Stadelmann.[47] Multislice calculations were performed using four HR-TEMs including: a JEOL 2010 LaB6 with a UHR or HRP pole piece, a CM 300 SuperTwin, a JEOL 2100-non $C_s$-corrected TEM, and a Titan 80-300. Crystallographic Information Files were created using Accelrys' Materials Studio and then imported to JEMS for image simulation through a range of defocus values near Scherzer defocus for each TEM. FFTs were applied to the simulated images to generate the {101}, {200}, and {002} plane families. We used image integration techniques to measure spot intensities in the FFTs.

*X-ray photoelectron spectroscopy*. XPS data was taken with a Kratos Axis Ultra DLD spectrometer with a monochromatic Al Kα x-ray source. High-resolution elemental scans were taken with a pass energy of 20 eV. Although a charge neutralizer was used for the measurement of bulk black phosphorus, charging inconsistencies of the C 1s peak prevented correction of the elemental P 2p peak to the C 1s peak. The 2D samples showed much less charging than the bulk samples, and energy correction with the C 1s peak was possible. Unless otherwise noted, all preparation was performed inside a nitrogen-filled glove box and transported to the XPS chamber in custom glassware. All sample transfers into/from the XPS chamber utilized a nitrogen-purged glove bag to prevent pre-mature oxidation by air and/or water vapor. The XPS loading chamber was vented with $N_2$. The black phosphorus samples (bulk) were mechanically cleaved using carbon or copper double-sided tape to expose a fresh (non-oxidized) surface; the tape was also used as a substrate for these experiments. Samples were subsequently removed from the XPS chamber and exposed to ambient air and room light for controlled periods of time. After air exposure, samples were loaded into the XPS and a phosphorus oxide peak was observed. The cycle was repeated over several iterations and growth of the oxide peak was quantitatively measured in relation to the elemental P 2p peak.

*UV-vis-nIR spectroscopy*. All spectroscopy was performed on a Cary 5000 double beam spectrometer with an external integrating sphere apparatus. Starna 1-mm pathlength quartz cuvettes with transparency range from 170 to 2700 nm were used in all experiments. Cuvettes were filled in a glove box and fitted with airtight PTFE stoppers to maintain an inert atmosphere. See Supporting Information, section 10 for a more detailed description. Concentration of suspended flakes was determined by correlating UV-vis data with inductively-coupled plasma mass spectrometry. Samples were prepared in NMP, centrifuged at 3,000 *g* and dialyzed into fresh NMP under inert conditions to remove possible molecular phosphorus byproducts caused by sonication (Millipore Biotech RC membranes, 8-10 kDa). See Supporting Information for more details.

ASSOCIATED CONTENT

**Supporting Information**. Full description of synthesis, characterization, and modeling. This material is available free of charge *via* the Internet at http://pubs.acs.org.

AUTHOR INFORMATION

**Corresponding Author**

*sw@unc.edu.

**Author Contributions**

The manuscript was written through contributions of all authors. All authors have given approval to the final version of the manuscript.

**Funding Sources**

S.C.W. acknowledges support of this research by UNC Chapel Hill startup funds, the UNC Office of Technology Development, and the National Science Foundation under grant DMR-1429407. T.W.F. and A.H.W. acknowledge support of this work by the National Science Foundation Graduate Research Fellowship under Grant No. DGE-1144081. Any opinion, findings, and conclusions or recommendations expressed in this material are those of the authors(s) and do not necessarily reflect the views of the National Science Foundation.

ACKNOWLEDGMENT

We thank Bruker Instruments for access to a FastScan AFM. We thank J. F. Cahoon and group members for access to furnaces, sonicators, and other equipment while our laboratory was built. We thank A. S. Kumbhar for assistance in HR-TEM imaging. We acknowledge P. Stadelmann, E. Samulski, J. DeSimone, J. F. Cahoon, K. L. Kuntz, D. Druffel, & S. Kolavennu for fruitful discussions. We thank T. Nilges and M. Köpf for sharing the details of their black phosphorus synthesis[46] prior to publication. A portion of this work was performed in the UNC-EFRC Instrumentation Facility established by the UNC EFRC (Solar Fuels and Next Generation Photovoltaics, an Energy Frontier Research Center funded by the U.S. Department of Energy, Office of Science, Office of Basic Energy Sciences under Award Number DE-SC0001011) and the UNC SERC ("Solar Energy Research Center Instrumentation Facility" funded by the US Department of Energy – Office of Energy Efficiency & Renewable Energy under Award Number DE-EE0003188).

TOC Graphic:

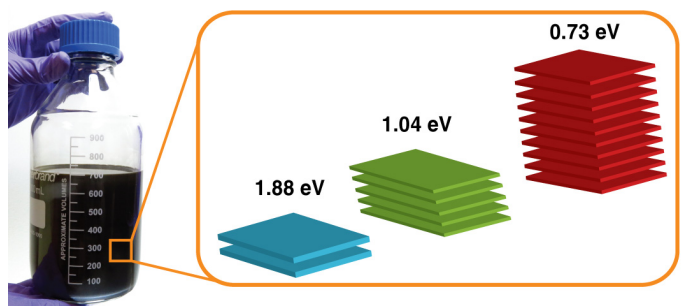

# Supporting Information

# Phosphorene: Synthesis, Scale-up, and Quantitative Optical Spectroscopy

*Adam H. Woomer[§], Tyler W. Farnsworth[§], Jun Hu[§], Rebekah A. Wells[§], Carrie L. Donley[‡], and Scott C. Warren[§†]*

[§]Department of Chemistry, [‡]Chapel Hill Analytical and Nanofabrication Laboratory, and [†]Department of Applied Physical Sciences, University of North Carolina at Chapel Hill, Chapel Hill, NC 27599, USA.

*Email: sw@unc.edu.

## Contents

## 1. Materials, supplies, and general notes on sample handling

All solvents were purchased from Sigma Aldrich unless otherwise noted. The solvents were degassed prior to use and handled under nitrogen. Black phosphorus (smart elements, ~99.998%), red phosphorus (Sigma Aldrich, ≥ 99.999+%), *N*-methyl-2-pyrrolidone (Sigma Aldrich or Acros Organics, anhydrous, 99.5%) 2-propanol (FisherSci, electronic grade), cyclopentanone (≥ 99%), 1-cyclohexyl-2-pyrrolidone (≥ 99%), 1-dodecyl-2-pyrrolidinone (≥ 99%), benzyl benzoate (≥ 99.0%), 1-octyl-2-pyrrolidone (≥ 98%), 1-vinyl-2-pyrrolidinone (≥ 99%), benzyl ether (≥ 98%), 1,3-dimethyl-2-imidazolidinone (≥ 99%), cyclohexanone (≥ 99.8%), chlorobenzene (≥ 99.5%), dimethylsulfoxide (≥ 99.9%), benzonitrile (anhydrous, ≥ 99%), *N*-methylformamide (≥ 99%), dimethylformamide (≥ 99%), benzaldehyde (≥ 99%).

All samples were prepared in a nitrogen-filled glove box ($O_2$ < 1 ppm) and transferred from instrument to instrument using custom-designed glassware with Rotaflo stopcocks. When necessary, instrument loading chambers and/or the entire instrument was enclosed in $N_2$-filled glove bags to provide an inert atmosphere.

## 2. Black phosphorus synthesis

In initial experiments, we used black phosphorus from smart elements. These samples were purchased on several occasions between December 2012 and April 2014. We characterized this material by $^{31}$P NMR, powder x-ray diffraction, and x-ray photoelectron spectroscopy. These studies confirmed that the material was black phosphorus. XPS revealed, however, that tin iodide was present in some samples, indicating that smart elements used the Nilges method of black phosphorus production[1]. In the summer of 2014, smart elements changed their synthetic method; we cannot comment on the quality of this new material.

Much of our work was carried out using home-grown black phosphorus, produced by sublimation of red phosphorus in the presence of Sn and $SnI_4$. The method is nearly identical to the newly published method of Nilges[2]. Red phosphorus (0.5 g), tin (10 mg), and tin iodide (15 mg) were loaded into a partially sealed quartz tube in a glove box. The tube was 14-cm long, had a 1-cm inner diameter, and a 1.4-cm outer diameter). The tube and its contents were evacuated to ca. $4 \times 10^{-3}$ mbar and sealed in a fume hood with an oxy-hydrogen torch while maintaining vacuum within the ampoule. We ensured that the ends of the quartz tubes were thick (> 0.4 cm): in one experiment, the quartz tube burst because the walls were too thin and because the pressure during the reaction exceeds atmospheric pressure. The reaction was carried out in a Lindberg Blue three-zone furnace using the temperature profile described by Nilges. We maintained a 60 °C temperature differential across the 15-cm tube by inserting insulation between the two zones. Black phosphorus crystals were opened in our glove box, soaked in anhydrous acetone, *N*-methyl-2-pyrrolidone, and isopropanol, and subsequently dried prior to use.

## 3. Atomic force microscopy

To assess the effectiveness of mechanical exfoliation as a method for producing monolayer 2D phosphorus, we used an atomic force microscope (AFM) to characterize flake thickness. Black phosphorus crystals were cleaved multiple (*ca.* 9 times) with scotch tape and then pressed onto quartz or silicon wafers with a 300 nm thermal oxide. Wafers were placed in acetone and then isopropanol for 10 minutes each to remove tape residue. We analyzed flake thickness using two atomic force microscopes: a Bruker Dimension FastScan and a Bruker Dimension Icon, both of which were equipped with motorized, programmable stages. To protect the samples from oxidation, FastScan measurements were performed under flowing nitrogen gas while Dimension Icon measurements were performed in a glove box. Scan areas were $15 \times 15$ μm$^2$ (FastScan) and $25 \times 25$ μm$^2$ (Dimension Icon). A script was used to automate image acquisition, thereby eliminating selection bias in these experiments. The acquisition time per image was approximately 1 minute (FastScan) and 20 minutes (Dimesion Icon). In total, we collected and analyzed 594 AFM images and 3,538 flakes. We found that the overall yield of thin phosphorus flakes (<5 nm) was 0.06%.

## 4. Method of liquid exfoliation:

Black phosphorus crystals were slightly crushed using a mortar and pestle in a nitrogen glove box. For typical experiments, 10 mg was weighed into a 20-mL scintillation vial. Twenty milliliters of solvent was added to give a concentration of 0.5 mg/mL. Vials were tightly capped and wrapped with parafilm to prevent air exposure before placing into a Branson 5800 bath sonicator. The bath sonicator was outfitted with a test tube rack to allow for controlled placement of vials. Vials were systematically moved through several locations during the course of sonication to minimize vial-to-vial variations in phosphorus dispersion. The samples were subjected to eight to ten cycles of sonication, each lasting 99 minutes. Bath water was changed after each cycle to maintain a temperature between 22 and 30 °C (during sonication, bath temperature increased dramatically). During the sonication process, the black phosphorus crystals dispersed into the solution and the suspension acquired a brown appearance. After sonication, the vials were returned to the glove box.

To fractionate phosphorus suspensions and isolate narrow thickness distributions of 2D phosphorus, we employed a three-step centrifugation protocol. First, solutions were transferred to Nalgene Oak Ridge FEP 10- or 50-mL centrifuge tubes. The solutions were centrifuged in a Sorvall RC-5B superspeed refrigerated centrifuge (rotor radius 10.7 cm). Second, the supernatant from the centrifuge tubes was collected and transferred to a clean centrifuge tube. Third, the supernatant was centrifuged at a speed higher than the first run. The sediment was collected and typically redispersed in fresh solvent. Depending on choice of centrifugation speeds, these fractionated suspensions contained 2D phosphorus with narrow and systematically varying thicknesses distributions (see Fig. 4 a,b).

As a typical example, a distribution could be collected at RCF values between 17,200$g$ and 23,400$g$. The tube would first be spun at 17,200$g$ for 30 minutes. The resulting supernatant would then be removed and re-centrifuged at 23,400$g$ for 30 minutes. The sediment from the second centrifuge would then contain a distribution of sheets that could then be redispersed into any solvent (often we chose IPA) for further analysis; for simplicity, we label this new suspension as 20,200$g$, the average RCF between the two sequential centrifugation steps. Table S1 lists the average RCF value for each suspension in Figure 4a. Note that all solution transfers between centrifuge tubes were performed inside a glove box. High speed centrifugations (>12,000$g$) were performed at 4 °C to lengthen tube lifetime.

**Table S1.** Labeling of centrifugation fractions containing 2D phosphorus suspensions

| Sequential RCF (g) | Label (1,000g) |
|---|---|
| 30 to 480 | 0.12 |
| 480 to 1,900 | 1.1 |
| 1,900 to 4,300 | 3.0 |
| 7,700 to 12,000 | 9.7 |
| 17,200 to 23,400 | 20.2 |

## 5. *Results of liquid exfoliation:*

Inductively coupled-mass spectroscopy (ICP-MS) was correlated with UV-vis-nIR spectroscopy to make a standard curve that we used to measure the concentration of liquid-exfoliated black phosphorus. Black phosphorus was exfoliated in NMP, centrifuged at 3,000$g$, and dialysed into fresh NMP under inert conditions to remove possible molecular phosphorus byproducts caused by sonication (Millipore Biotech RC membranes, 8-10 kDa). The dialyzed samples were serially diluted and analyzed by ICP-MS and UV-vis-nIR spectroscopy at 450, 500, and 550 nm to create three calibration curves. The equations governing the relationship between absorbance (see section 9 for a description of UV-vis-nIR methods) for 2D phosphorus supernatants collected after centrifugation at 3,000$g$ were:

$A_{450\text{-nm}} = 3.4$ E-6 × (phosphorus concentration in parts per billion) [1]
$A_{500\text{-nm}} = 2.5$ E-6 × (phosphorus concentration in parts per billion) [2]
$A_{550\text{-nm}} = 1.8$ E-6 × (phosphorus concentration in parts per billion). [3]

Table S2 summarizes the results of our liquid exfoliation study.

**Table S2.** Liquid exfoliation of black phosphorus and Hansen solubility parameters[3].

| Solvent | Average Conc. (ug/mL) | Std. Dev. (ug/mL) | Hildebrand (MPa$^{1/2}$) $\delta$ | Hansen (MPa$^{1/2}$) $\delta_d$ | $\delta_p$ | $\delta_h$ |
|---|---|---|---|---|---|---|
| N-methyl-2-pyrrolidone (NMP) | 44.12 | 11.23 | 23 | 18 | 12.3 | 7.2 |
| Cyclopentanone | 37.05 | 17.30 | 22.1 | 17.9 | 11.9 | 5.2 |
| 1-Cyclohexyl-2-pyrrolidone (CHP) | 25.08 | 7.54 | 20.5 | 18.2 | 6.8 | 6.5 |
| 1-Dodecyl-2-pyrrolidinone (N12P) | 22.81 | 6.26 | 18.3 | 17.5 | 4.1 | 3.2 |
| Benzyl benzoate | 32.05 | 16.69 | 21.3 | 20 | 5.1 | 5.2 |
| 1-Octyl-2-pyrrolidone (N8P) | 37.47 | 15.72 | 19.1 | 17.4 | 6.2 | 4.8 |
| 1-Vinyl-2-pyrrolidinone (NVP) | 61.74 | 20.75 | 19.8 | 16.4 | 9.3 | 5.9 |
| Benzyl ether | 3.31 | 3.88 | 20.6 | 19.6 | 3.4 | 5.2 |
| 1,3-Dimethyl-2-imidazolidinone | 65.63 | 10.20 | 23 | 18 | 10.5 | 9.7 |
| Cyclohexanone | 3.54 | 2.28 | 20.3 | 17.8 | 8.4 | 5.1 |
| Chlorobenzene | 0.76 | 1.02 | 19.6 | 19 | 4.3 | 2 |
| Dimethylsulfoxide (DMSO) | 29.06 | 2.19 | 26.7 | 18.4 | 16.4 | 10.2 |
| Benzonitrile | 110.82 | 16.63 | 22.5 | 18.8 | 12 | 3.3 |
| N-methylformamide | 50.91 | 17.61 | 30.1 | 17.4 | 18.8 | 15.9 |
| Dimethylformamide | 40.82 | 4.72 | 24.9 | 17.4 | 13.7 | 11.3 |
| Benzaldehyde | 16.12 | 13.52 | 21.4 | 19.4 | 7.4 | 5.3 |
| Isopropylalcohol (IPA) | 38.65 | 7.60 | 23.6 | 15.8 | 6.1 | 16.4 |

## *6. Transmission electron microscopy*

To analyze and characterize phosphorus flakes, a JEOL 100CX II TEM was used for low resolution imaging. The TEM was operated at 100 kV accelerating voltage and had a resolution of 2 Å (lattice) and 3 Å (point to point). For HR-TEM, we used a JEOL 2010F-FasTEM at 200kV accelerating voltage. The resolution for this instrument was 1 Å (lattice) and 2.3 Å (point to point). In a glove box, phosphorus suspended in IPA or NMP was dropcast (~10 µL) onto a 300-mesh copper TEM grid coated with Lacey carbon (Ted Pella). After deposition, the solvent was evaporated for one hour (IPA) or at least 48 hours (NMP). During this period, TEM grids were held with self-closing tweezers and were not blotted against filter paper to ensure that a representative distribution of the material was deposited. The TEM grids were placed in a grid box wrapped in aluminum foil, placed in a nitrogen-filled zip-loc bag, and brought to the TEM. Grids were rapidly transferred into the TEM with a minimal amount of light exposure. For both TEM analyses, images were acquired and processed with Gatan Digital Micrograph. Fast Fourier Transforms (FFTs) were applied to HR-TEM images to observe the plane families present in the HR-TEM images of phosphorus flakes.

## 7. *Edge contrast analysis*

To create the histograms of thickness and lateral dimensions, the line profilometer tool of Gatan Digital Micrograph was used. We report the lateral dimension of a phosphorus flake as the diagonal of two perpendicular length measurements. We here present a full method for our edge contrast analysis approach to determine thickness:

i. *Alignment* – TEM samples are first prepared and loaded into a JEOL 100CX II TEM as described in Section 6 above. The microscope was carefully aligned, including objective lens, condenser lens, gun position, eucentric height and sample tilt. The latter two alignments are of critical importance to ensure there is no change in defocus or variation in sample height when translating across the grid.

ii. *Imaging* – A charge-coupled device (CCD) camera was used to acquire all TEM images. The scintillator crystal and camera were pristine and well maintained, and thus there were no artifacts or variation in background seen in the resulting images. Each image was taken at identical imaging conditions including exposure time, magnification, and defocus value. All images were taken at Scherzer defocus with a complete refocusing of diffracted electrons, and only minor, if any, defocus adjustments were necessary between samples. No pictures were analyzed near the edge of the grid bars, due to the artificial uniform increase in brightness.

iii. *Contrast Analysis* - Contrast changes across phosphorus flake edges were measured by drawing a line profile across and perpendicular to the edge of the flake (Fig. 3.a, line and inset). To first establish the thickness-contrast relationship for our TEM, we measured contrast change for 500 flakes and found that the smallest intensity change was $25 \pm 3$ counts and all other intensity changes were multiples of 25 counts (Fig. 3.c, inset). We therefore assigned an intensity change of 25, 50, 75, and 100 counts to monolayers, bilayers, trilayers, and four-layered 2D phosphorus flakes, respectively. We note that in some cases, the edge of the flake was over vacuum, and in other cases it was over a carbon film. The maximum difference between the two measurements was 9 counts, and the average difference was $4.4 \pm 3.0$ counts (N = 12). These numbers are considerably smaller than the average counts per layer of $25 \pm 3$ counts. We conclude that there was, in general, a negligible difference between these two cases.

iv. *Sample Selection* – The analysis of edge contrast is central for the accurate determination of the optical properties of 2D phosphorus, however it also has the highest potential to introduce an analytical bias. We have identified the five most common occurrences for bias in the contrast analysis and here report their prevalence or prevention.

   a. Organic Residue: Solvents and environmental organic contaminants will vaporize into an amorphous carbon film if not properly removed from the TEM grid, contributing a non-uniform contrast to the flake and lacey carbon film. To account for this, we evaporated our solvent for extended periods of time (30 minutes to 48 hours) under vacuum. Samples with a non-uniform contrast due to organics were discarded from our analysis. This error is independent of flake thickness and we find that *c.a.* 10% of samples across all three suspensions were affected by organic residue. We note that thermal degassing could further eliminate organic residue.

b. Polycrystallinity: Phosphorus flakes with a large degree of polycrystallinity will always exhibit some amount of diffraction contrast that is not uniform across the flake, no matter how well the microscope is aligned or focused. We removed these flakes with a low speed centrifugation (500 rpm), and any flakes imaged showing diffraction contrast were not included in our final analysis. These flakes were *c.a.* 10% of all three combined phosphorus suspensions.

c. Multilayer flakes: The presence of flakes with more than one distinct thickness is uncontrollable, however this was only somewhat common in the thickest flake distributions (*c.a.* 15% of flakes have more than one thickness). These flakes are predominantly of one thickness, with >60% of flake area having the corresponding uniform contrast, and were thus assigned to this value for our optical analysis. In the medium and thinner distributions, nearly all flakes exhibited a complete uniform contrast. The overall representation of multilayer flakes for the combined three fractions was *c.a.* 6%.

d. Aggregation: During our TEM analysis, we have noticed a tendency for flakes to aggregate. We found that this occurrence is related to the volatility of the liquid exfoliation solvent as well as the deposition technique. To prevent this we used isopropanol as a high vapor pressure solvent for TEM samples prepared specifically for edge contrast analysis. Because of the complexity of a 2D phosphorus aggregated structure, flakes can appear to be amorphous, polycrystalline, and have non-uniform contrast. We have chosen to not analyze any flakes that are a component of an aggregate.

e. Lateral Size: An important criteria for the proper execution of the edge contrast thickness measurements is constant magnification, however this can complicate the analysis of small 2D phosphorus flakes (<15 nm linear dimension for our chosen magnification). We have sought to analyze all flakes with size larger than 15 nm, and the observation of flakes smaller than this size was rare, (c.a. <1%). After applying the following techniques to eliminate sample bias, we believe we have accurately accounted for a representative sample of 2D phosphorus flakes in each suspension.

## 8. *Multislice Calculations*

To simulate HR-TEM images of 2D phosphorus from one to eight layers, we used Java electron microscopy software (JEMS) version 4.1830U2014 by Pierre Stadelmann[4]. Multislice calculations were performed using four HR-TEMs including: a JEOL 2010 LaB6 with a UHR or HRP pole piece, a CM 300 SuperTwin, a JEOL 2100-non Cs corrected TEM, and a Titan 80-300. Crystallographic Information Files were created using Accelrys' Materials Studio and then imported to JEMS for image simulation through a range of defocus values near Scherzer defocus for each TEM. FFTs were applied to the simulated images to generate the {101}, {200}, and {002} plane families. We then used image integration techniques to measure spot intensities in the FFTs. From this analysis, we found that for an even-layered phosphorus flake, the 101 diffraction spot is absent in all simulations, regardless of microscope or defocus value. In all odd-layered cases other than a monolayer, the 101 intensity is severly diminished at or near Scherzer defocus (Fig. 3f). Table S3 summarizes the imaging conditions and identifies the defocus range at which this trend is true for each microscope.

**Table S3**. Scherzer defoci and defocus range for a large 101:200 ratio for monolayer 2D phosphorus

| Microscope | Accelerating Voltage (kV) | $C_S$ (mm) | Scherzer Defocus (nm) | Defocus Range (nm) |
|---|---|---|---|---|
| CM 300 ST | 300 | 1.20 | 59.0 | 60 to 62 |
| Jeol 2010F - HRP | 200 | 1.00 | 61.0 | 64 to 68 |
| Jeol 2010F - UHR | 200 | 0.50 | 43.4 | 40 to 44 |
| Jeol 2100 | 200 | 1.40 | 72.0 | 68 to 71 |
| Titan 80-300 | 300 | -0.03 | -9.0 | -11 to -13 |

## 9. *X-ray photoelectron spectroscopy*

XPS data was taken with a Kratos Axis Ultra DLD spectrometer with a monochromatic Al Kα x-ray source. High-resolution elemental scans were taken with a pass energy of 20 eV. Although a charge neutralizer was used for the measurement of bulk black phosphorus, charging inconsistencies of the C 1s peak prevented correction of the elemental P 2p peak to the C 1s peak. The 2D samples showed much less charging than the bulk samples, and energy correction with the C 1s peak was possible.

Unless otherwise noted, all preparation was performed inside a nitrogen-filled glove box and all sample transfers into/from the XPS chamber utilized a nitrogen-purged glove bag to prevent air exposure. Custom glassware was used to transport the XPS sample bar to and from the glove box and the instrument, thus preventing premature oxidation of the sample by air or water vapor. The XPS chamber was vented with $N_2$.

Black phosphorus samples (bulk) were mechanically cleaved using carbon or copper tape to expose a fresh (non-oxidized) surface. Initial high resolution XPS scans of the P 2p peak showed no signs of phosphorus surface oxidation. Samples were subsequently removed from the XPS chamber and exposed to ambient air and room light for a set period of time. After air exposure, samples were loaded into the XPS and a phosphorus oxide peak was observed. The cycle was repeated over several iterations and growth of the oxide peak was quantitatively measured in relation to the elemental P 2p peak.

XPS was also used to probe the surface oxidation of few-layer phosphorus sheets prepared by liquid exfoliation. In a glove box, the sheets were doctor-bladed into thin films on Au-plated silicon wafers. After measuring the initial spectrum, the sample and its holder were placed into a custom transport glassware filled with industrial grade oxygen and illuminated with a 460 nm LED for 1 hour. XPS analysis confirmed the presence of a phosphorus oxide peak.

## *10.2D phosphorus UV-vis-nIR absorbance spectroscopy*

Starna 1-mm pathlength quartz cuvettes with transparency range from 170 to 2700 nm were used in all experiments. Cuvettes were filled in a glove box and fitted with airtight PTFE stoppers to maintain an inert atmosphere.

In UV-vis-nIR transmission spectroscopy, the incident light (I) may be transmitted (T), absorbed (A), reflected (R), forward-scattered (FS) or back-scattered (BS). Because of the conservation of energy, $I = T + A + R + FS + BS$, where each term measures light intensity. In the solution spectroscopy of small molecules, typical approximations are that $FS \approx 0$, $BS \approx 0$, and that R is made effectively zero *via* the collection of a background spectrum that contains solvent but no analyte. Consequently, $I = T + A$, allowing a simple measurement of the transmitted light intensity with (T) and without (I) the analyte to provide accurate information about light absorption (A) by the molecule.

These approximations do not hold when performing transmission spectroscopy on objects that are larger than a few nanometers in size. As objects increase in size, so does the scattering cross-section. Objects with linear dimensions greater than *ca.* $\lambda/15$ will scatter more light in the forward direction than in the back direction due to the deconstructive interference of the back-scattered light[5]. Noting this effect, we sought to quantify FS light, which could be achieved in a straightforward way (see below), to estimate whether the BS light, which is harder to directly measure, could be neglected.

To capture forward-scattered light, we made four experimental choices, summarized in Figure S1. First, the cuvette was placed near the detector opening. This allowed the collection of light scattered at high angles. Second, we selected a very thin (1-mm thick) cuvette rather than a 1-cm cuvette. By having the entire solution placed close to the detector opening, we could collect light scattered at very high angles. Third, by ensuring that the light that illuminated the cuvette covered a region that was smaller than the detector's entrance aperture, we were able to collect an even larger proportion of forward-scattered light. Fourth, the detector was an integrating sphere, which allowed the detector to collect and quantify light, even if it was scattered at a high angle.

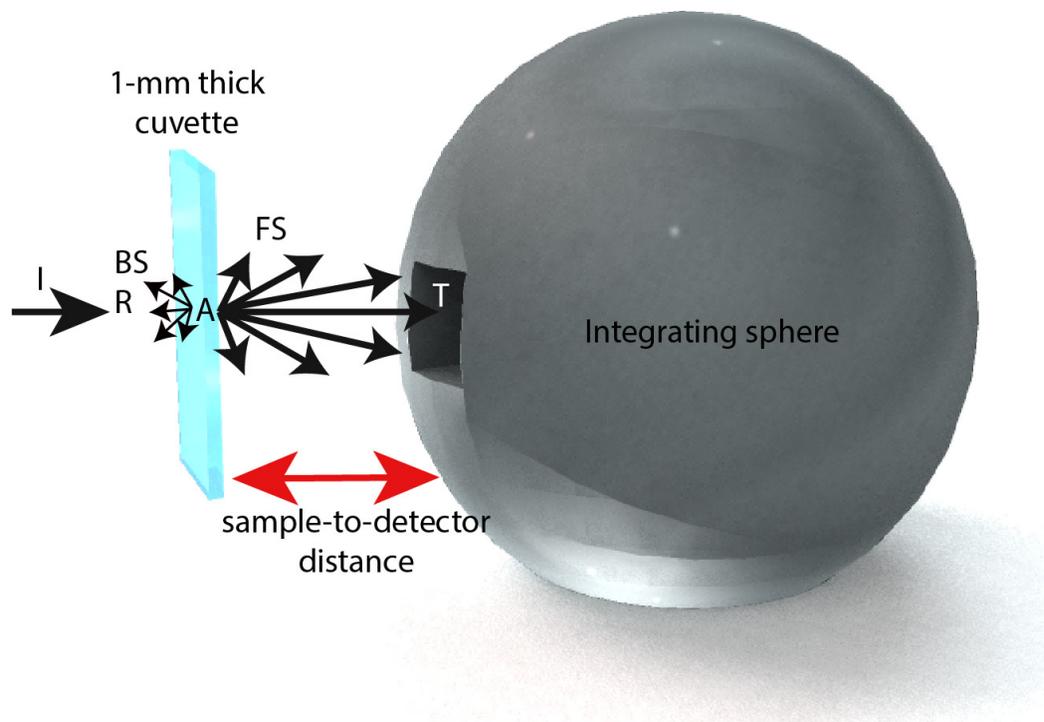

**Figure S1.** UV-vis-nIR absorbance spectroscopy setup. By placing the 1-mm cuvette adjacent to the integrating sphere aperture and by restricting the size of the incident light beam I, we were able to capture most forward-scattered (FS) light.

In order to test the utility of our design, we systematically varied the distance between the sample and detector, collecting "absorbance" spectra at each position. The solution used for this test was comprised of thick and large flakes, which scatter a larger proportion of light than in most of our samples. Figure S2 summarizes these results. In moving the sample from a position nominally labeled 4 cm (the actual sample-to-detector distance is *ca.* 4.5 cm) to 0 cm (the actual sample-to-detector distance is *ca.* 0.5 cm), the absorbance spectra changed monotonically, resulting in an apparent decrease in absorbance. We further modified the setup to remove the stand-alone holder that held the sample in place, which allowed us to clamp the sample directly against the opening aperture of the integrating sphere (sample-to-detector distance is *ca.* 0.15 cm). By systematically decreasing the sample-to-detector distance, we captured FS light at angles between 13° and 147°. As the angle of detection increased, we observed a decrease in apparent absorbance that was consistent with capturing increased amounts of forward-scattered light. These changes in apparent absorbance resulted in quantifiable shifts in Tauc plot analyses which we then used to calculate the error associated with assuming BS $\approx$ 0. From our angle-dependent measurements, we estimate that neglecting BS light could lead to, at most, a 3% underestimate of the band gap for a particular thickness of 2D phosphorus. Because of the relatively small error, we used this modified setup with the sample held in the "clamp" position to collect all of our UV-vis-nIR transmission spectra. An important conclusion of these experiments is that the total amount of scattered light (both FS and BS) is relatively small when compared to the amount of light that is absorbed.

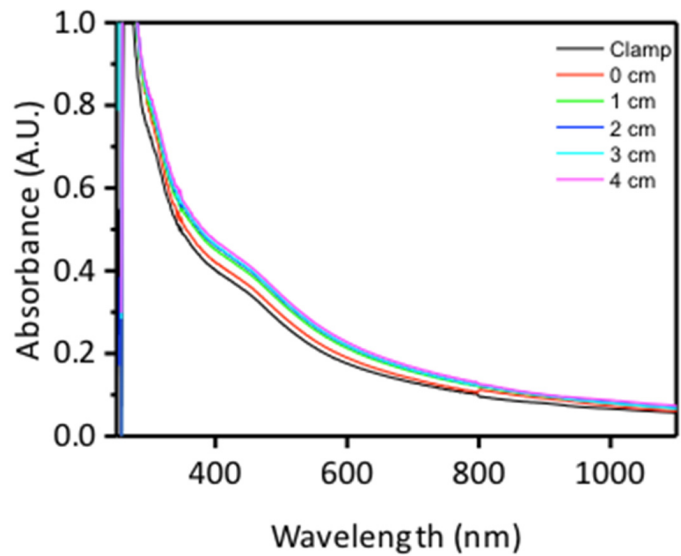

**Figure S2.** Apparent absorbance vs. wavelength for different cuvette positions.

## 11. Exciton Binding Energy

We determined the exciton binding energy (EBE) of black phosphorus for varying dielectric constants by extrapolating a plot given by Rodin and coworkers[6]. We determined a best fit function for their data by using the Matlab Curve Fitting Toolbox. We found that a rational function of order zero for the numerator and order two for the denominator gave the best fit for the data, show in Figure S3. Plotting this function allowed us to determine the EBE for solvents characterized by different dielectric constants. For *N*-methyl-2-pyrrolidone ($\kappa = 32.17$) and isopropanol ($\kappa = 20.18$) the calculated EBE of black phosphorus is -15 meV and -32 meV, respectively.

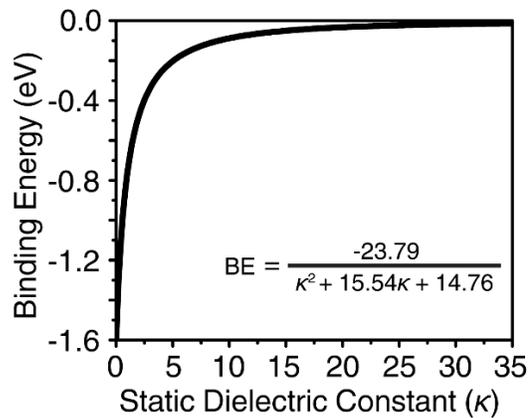

**Figure S3.** Exciton binding energy for different dielectric constants.

## 12. Band gap measurement of 2D phosphorus solutions

In order to validate the alpha method, we used calculated $G_0W_0$ absorption spectra[7] (Figure S4) of bulk, trilayer, bilayer, and monolayer phosphorus. The band gaps were calculated as the minimum energy difference between the valence and conduction bands of these calculated band structures and are referred to as "reported band gaps" throughout.

We next applied our alpha method to the same absorption spectra to obtain a new estimate of the band gap. To do this, we measured the absorption coefficient at the theoretical bulk band gap (0.3 eV) and equated this value to $\alpha_{AE}$. Drawing a line across the plot from the standard $\alpha_{AE}$ allowed us to determine a band gap for each flake thickness (Figure S4). We directly compare the alpha estimates to the reported band gaps in Table S4. As listed in Table S4, the percent difference between our alpha method of extracting the band gap and the reported band gaps remains relatively small, with a maximum difference of 1.85%. We conclude that the alpha method is a reliable measure of band gap for samples that contain a single flake thickness.

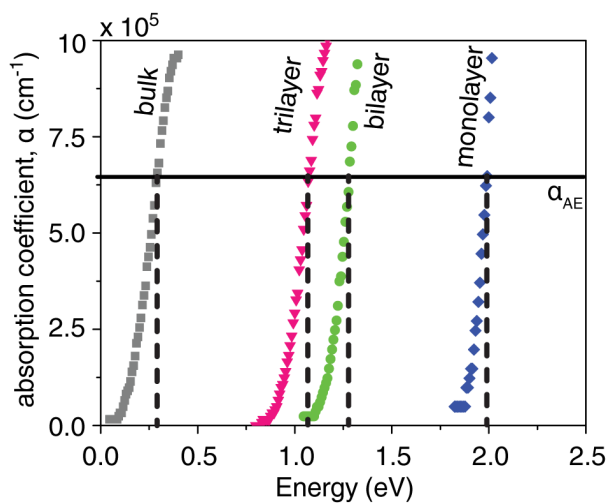

**Figure S4.** Alpha method of calculated $G_0W_0$ absorption spectra[7]

**Table S4.** Comparison of band gaps extracted using both the reported band gap and alpha method

|  | Reported band gap (eV) | $\alpha = 6.98 \times 10^5$ cm$^{-1}$ (eV) | Percent Error (%) |
|---|---|---|---|
| **Bulk** | 0.30 | 0.30 | n/a |
| **Trilayer** | 1.06 | 1.08 | 1.85 |
| **Bilayer** | 1.30 | 1.283 | 1.33 |
| **Monolayer** | 2.00 | 1.995 | 0.25 |

Our samples, however, are polydisperse. In order to account for their polydispersity, we first considered two different statistical measures—number- and weight-averaged thickness (these are the 1st and 2nd moments of a distribution and are analogous to $M_n$ and $M_w$ in polymer chemistry)—to estimate an "effective" flake thickness to best describe a polydisperse distribution. If we consider a simple mixture of 10% bilayers and 90% trilayers and calculate the number-averaged thickness:

$$number\ averaged\ thickness = \sum_t x_t t = 0.1 \times 2 + 0.9 \times 3 = 2.90$$

where $x_t$ is the number fraction of flakes with a specific thickness and $t$ is the thickness. The weight-averaged thickness for the same mixture is:

$$weight\ averaged\ thickness = \sum_t w_t t = 0.07 \times 2 + 0.93 \times 3 = 2.93$$

where $w_t$ is the weight fraction of flakes with a specific thickness and is calculated by

$$w_t = \frac{x_t t}{\sum_t x_t t}$$

which, in the case of the 10:90 bilayer:trilayer mixture, is:

$$w_2 = \frac{0.1 \times 2}{0.9 \times 3 + 0.1 \times 2} = 0.07$$

and

$$w_3 = \frac{0.9 \times 3}{0.9 \times 3 + 0.1 \times 2} = 0.93$$

Then we now have two measures of "effective thickness" to describe the bilayer:trilayer distribution, 2.90 and 2.93. If we were to apply our alpha method to the absorption spectra of the bilayer:trilayer mixture, we could correlate a measured "effective band gap" with these effective thicknesses from the statistical analysis. These effective thickness-effective band gap data points can be compared to the power law fit (described in main text) to provide an estimate of the error associated with the effective thickness-effective band gap approach.

To illustrate how these statistical approaches influence our experimental results, we step away from the calculations and examine how the number- and weight-average distributions describe our real polydisperse distributions. In some situations, a weight average is advantageous because the weighting applied for each flake thickness depends on how much of that flake is present, rather than the number of flakes of that type that are present. Figure S6 shows a number-average thickness distribution and Figure S7 shows a weight-average thickness distribution for our real flake histograms. As noted in the manuscript, the weight-averaged distribution is not equivalent to a weight fraction, but rather accounts for both the thickness and area of each flake. We will describe in more detail, below, which sorts of distributions (or, more specifically, statistical measures of the distribution) are best for pairing with the alpha method.

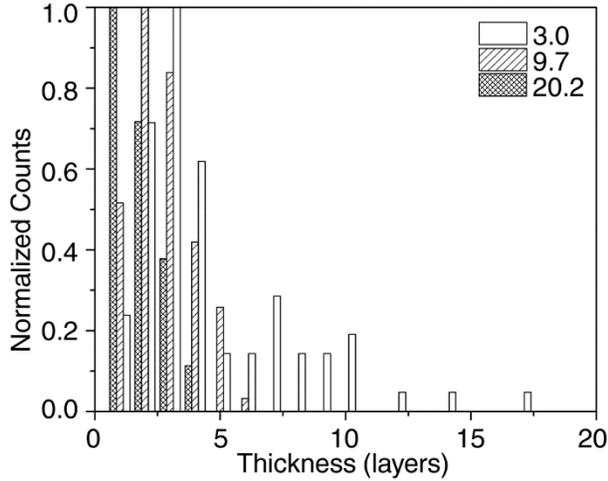

**Figure S6.** Number-average thickness distribution of three 2D phosphorus suspensions.

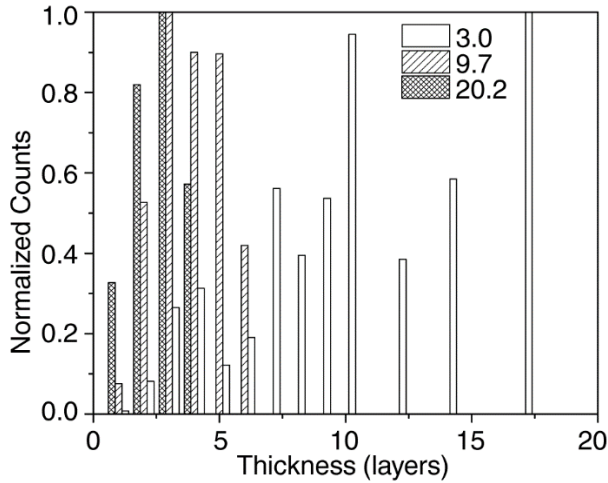

**Figure S7.** Weight-average thickness distribution of three 2D phosphorus suspensions.

In order to determine which statistical measure of thickness should be paired with the alpha analysis, we return to the $G_0W_0$ calculations and expand our survey of statistical measures beyond number- and weight-averages to include three log-normal fittings: the mean, median, and mode. We examined mixtures of flakes of four different thicknesses, from bilayer flakes to five-layer flakes. The absorbance spectra of the bilayer and trilayer flakes came from $G_0W_0$ calculations[7] and the spectra of the four- and five-layer were estimated by scissoring the band gap of the 3-layer flake until it had band gaps that matched those predicted by a power law. We note that the absorption coefficient (α) at energies near the band gap of the 3-layer flake is nearly the same as for the bulk material[7] so that scissoring to intermediate thicknesses is unlikely to introduce significant error in the variation of α with photon energy. The flakes were combined into 24 realistic distributions with systematically varying shapes and skewnesses (Figure S8). These distributions were used to weight the absorbance spectra prior to adding them to yield a total absorbance of each simulated mixture. The alpha method was applied to each mixture to obtain an effective band gap and five statistical measures were applied to analyze the distribution to obtain an effective thickness. A table that reports the most important parameters of these distributions is shown below, Table S6. This table is divided into several sections, including a listing of the number fraction of each flake in the distribution (blue), the effective band gap of the distribution as determined by the application of the alpha method (yellow), a listing of the effective thicknesses as determined by the five different statistical measures (orange), and the percent error between the "true" band gap—the band gap indicated by the power law—and the effective band gap, as determined by the alpha method. These comparisons between true and effective band gaps are performed five times for each distribution—one time for each of the five estimates of effective thickness. In other words, we use the power law equation to solve for the band gap at all five of the effective thicknesses to obtain five estimated band gaps. The difference in energy between this band gap, which falls on the power law curve, and the band gap that arises from the alpha method is used to calculate the percent error.

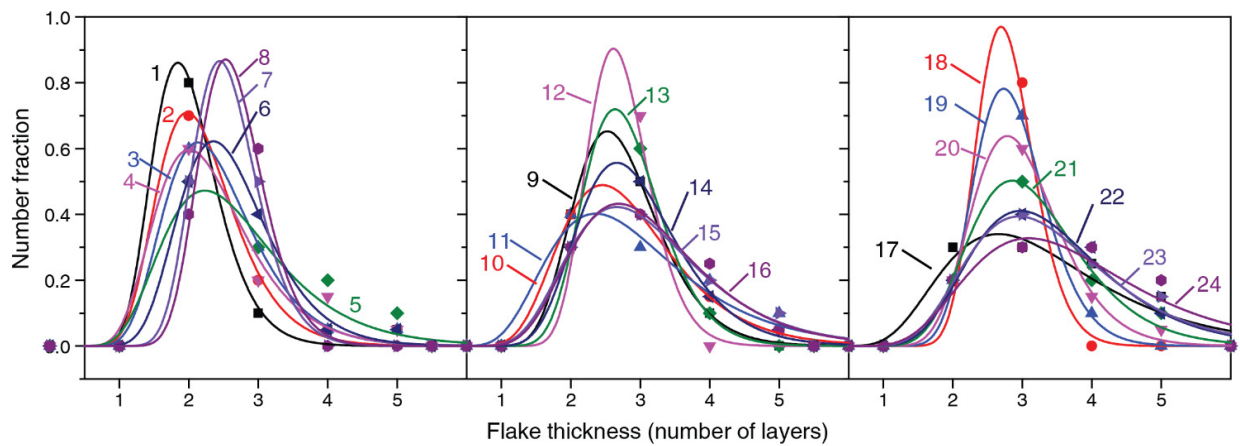

**Figure S8.** Summary of the 24 flake thickness distributions from Table S6. The distributions were fitted to a log-normal curve.

**Table S6.** Summary of 24 simulated mixtures. The individual spectra were obtained from calculated $G_0W_0$ spectra whose band gaps for 2- to 5-layer pieces fit a power law. The spectra were weighted according to their number fraction to determine an effective band gap and effective thickness. These values were compared to the value estimated from the power law fit to determine an error.

| Number fraction of each flake (2, 3, 4, or 5 layers thick) | | | | Effective band gap (eV) | Effective thickness of distribution | | | | | Error (% difference between power law band gap and effective (alpha) band gap) | | | | |
|---|---|---|---|---|---|---|---|---|---|---|---|---|---|---|
| | | | | | Log-normal fit | | | No fit | | Log-normal fitting | | | No fitting | |
| 2 | 3 | 4 | 5 | | Mean | Median | Mode | Number-average mean | Weight-average mean | Mean | Median | Mode | Number-average mean | Weight-average mean |
| 0.8 | 0.1 | 0.05 | 0.05 | 1.23 | 2.0 | 2.0 | 1.8 | 2.4 | 2.6 | 1.6% | 3.2% | 6.5% | -6.1% | -11.1% |
| 0.7 | 0.2 | 0.05 | 0.05 | 1.20 | 2.2 | 2.1 | 2.0 | 2.5 | 2.7 | -1.0% | 1.0% | 5.0% | -6.0% | -10.7% |
| 0.6 | 0.3 | 0.05 | 0.05 | 1.15 | 2.4 | 2.3 | 2.1 | 2.6 | 2.8 | -1.4% | 0.7% | 5.1% | -4.0% | -8.4% |
| 0.6 | 0.2 | 0.15 | 0.05 | 1.15 | 2.3 | 2.2 | 2.0 | 2.7 | 3.0 | 0.1% | 2.9% | 9.0% | -5.8% | -10.8% |
| 0.5 | 0.3 | 0.2 | 0.1 | 1.02 | 2.7 | 2.5 | 2.2 | 2.9 | 3.3 | 5.4% | 8.7% | 15.9% | 1.5% | -3.7% |
| 0.5 | 0.4 | 0.05 | 0.05 | 1.10 | 2.6 | 2.5 | 2.4 | 2.7 | 2.9 | -0.5% | 1.2% | 4.8% | -1.2% | -5.3% |
| 0.5 | 0.5 | 0 | 0 | 1.10 | 2.6 | 2.5 | 2.4 | 2.5 | 2.6 | -0.4% | 0.5% | 2.2% | 1.1% | -0.9% |
| 0.4 | 0.6 | 0 | 0 | 1.07 | 2.7 | 2.6 | 2.5 | 2.6 | 2.7 | 1.1% | 1.9% | 3.6% | 2.2% | 0.4% |
| 0.4 | 0.5 | 0.1 | 0 | 1.07 | 2.7 | 2.7 | 2.5 | 2.7 | 2.9 | -0.1% | 1.3% | 4.1% | 0.7% | -1.9% |
| 0.4 | 0.4 | 0.15 | 0.05 | 1.04 | 2.9 | 2.7 | 2.5 | 2.9 | 3.1 | 0.1% | 2.5% | 7.8% | 0.1% | -4.0% |
| 0.4 | 0.3 | 0.2 | 0.1 | 1.02 | 3.0 | 2.7 | 2.4 | 3.0 | 3.3 | 0.4% | 4.2% | 12.5% | -0.2% | -5.1% |
| 0.3 | 0.7 | 0 | 0 | 1.05 | 2.7 | 2.7 | 2.6 | 2.7 | 2.8 | 1.7% | 2.4% | 3.8% | 2.2% | 0.8% |
| 0.3 | 0.6 | 0.1 | 0 | 1.04 | 2.8 | 2.8 | 2.6 | 2.8 | 2.9 | 1.2% | 2.2% | 4.4% | 1.3% | -0.9% |
| 0.3 | 0.5 | 0.15 | 0.05 | 1.02 | 2.9 | 2.8 | 2.7 | 3.0 | 3.2 | 0.6% | 2.3% | 5.7% | 0.5% | -2.9% |
| 0.3 | 0.4 | 0.2 | 0.1 | 1.00 | 3.1 | 3.0 | 2.7 | 3.1 | 3.4 | -0.3% | 2.4% | 8.2% | 0.4% | -3.7% |
| 0.3 | 0.4 | 0.25 | 0.05 | 1.00 | 3.2 | 3.0 | 2.7 | 3.1 | 3.3 | -0.5% | 2.1% | 7.5% | 1.1% | -2.5% |
| 0.3 | 0.3 | 0.25 | 0.15 | 0.97 | 3.4 | 3.1 | 2.7 | 3.3 | 3.6 | -1.3% | 2.6% | 11.2% | 0.8% | -3.7% |
| 0.2 | 0.8 | 0 | 0 | 1.04 | 2.8 | 2.8 | 2.7 | 2.8 | 2.9 | 1.8% | 2.3% | 3.5% | 1.5% | 0.5% |
| 0.2 | 0.7 | 0.1 | 0 | 1.02 | 2.9 | 2.8 | 2.7 | 2.9 | 3.0 | 1.8% | 2.7% | 4.4% | 1.4% | -0.3% |
| 0.2 | 0.6 | 0.15 | 0.05 | 1.01 | 3.0 | 2.9 | 2.8 | 3.1 | 3.2 | 1.2% | 2.4% | 4.8% | 0.3% | -2.4% |
| 0.2 | 0.5 | 0.2 | 0.1 | 0.99 | 3.2 | 3.1 | 2.9 | 3.2 | 3.4 | 0.2% | 1.9% | 5.5% | -0.1% | -3.4% |
| 0.2 | 0.4 | 0.3 | 0.1 | 0.96 | 3.4 | 3.3 | 3.0 | 3.3 | 3.5 | -0.9% | 1.3% | 6.1% | 1.0% | -2.3% |
| 0.2 | 0.4 | 0.25 | 0.15 | 0.96 | 3.5 | 3.3 | 3.0 | 3.4 | 3.6 | -0.9% | 1.6% | 6.9% | 0.6% | -3.0% |
| 0.2 | 0.3 | 0.3 | 0.2 | 0.93 | 3.8 | 3.5 | 3.1 | 3.5 | 3.8 | -2.1% | 1.0% | 7.6% | 1.6% | -2.2% |

From this analysis of 24 distributions, we find that the log-normal mean and the number-average mean provide, on average, the best measures of flake thickness. It is these two measures that, when paired with the effective band gap, fall closest to the power law curve. The next best measures of effective thickness come from the log-normal median and weight-average mean. In every case, the log-normal median provides an effective thickness that is too small, while in almost every case, the weight-average mean provides an effective thickness that is too large. We therefore use the log-normal median and weight-average mean to define the likely range of the band gap in Figure 9a and we use the log-normal mean and number-average mean to define the range of most probable band gaps.

We note that the log-normal fitting allows us to determine the skewness of the distribution. Unfortunately, there is only a weak correlation between the distribution's skewness and the % error in the band gap, which prevents us from further refining our estimates of thickness and band gap by taking into account the distribution's shape (Figure S9).

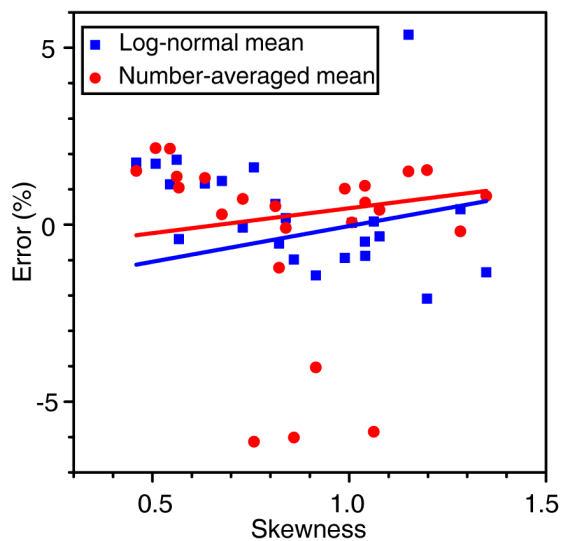

**Figure S9.** Correlation between the flake thickness distribution's skewness (determined from the log-normal curve fit, Figure S8) and % error between the true (power law) fit and the effective band gap-effective thickness estimate.

This analysis of artificial flake distributions has allowed us to identify the best statistical measures of flake thickness to pair with the alpha method. This information has allowed us to measure and accurately analyze real samples—obtaining their thickness distribution from TEM measurements and their absorption coefficients from UV-vis-near IR spectroscopy and ICP-MS—to establish a relationship between flake thickness and band gap (Figure 9). These curves are constructed by making a power law fit (described in main text) to the effective thicknesses and effective band gaps of real samples.

The power laws that are used to construct Figures 9 are shown below in Table S7. As mentioned previously, the best two fits to the data are the number-average mean and the log-normal mean and these form the boundaries of the darker regions in Figures 9. The weight-average mean and the log-normal median tend to over- and underestimate the band gaps, as described previously, and these define the boundaries of the lighter regions in Figures 9.

**Table S7.** Fitted curves of the thickess-dependent band gap analysis

|  | Low Energy Transition | High Energy Transition |
|---|---|---|
| **weight-average** | $E_g = \dfrac{2.64749}{N^{0.55131}} + 0.33$ | $E_g = \dfrac{3.58266}{N^{1.01981}} + 1.95$ |
| **number-average** | $E_g = \dfrac{2.49626}{N^{0.69007}} + 0.33$ | $E_g = \dfrac{2.95005}{N^{1.18685}} + 1.95$ |
| **log-normal mean** | $E_g = \dfrac{3.1542}{N^{1.02338}} + 0.33$ | $E_g = \dfrac{3.94145}{N^{1.63149}} + 1.95$ |
| **log-normal median** | $E_g = \dfrac{2.56742}{N^{0.96246}} + 0.33$ | $E_g = \dfrac{2.83518}{N^{1.53256}} + 1.95$ |

## *13. Supporting References*